\documentclass[11pt,a4paper]{article}
\usepackage{jheppub}
\usepackage{amsfonts,colordvi}
\usepackage{color}
\usepackage{amssymb}

\newcommand{\eq}{\begin{equation}}
\newcommand{\en}{\end{equation}}
\newcommand{\eqn}{\begin{eqnarray}}
\newcommand{\enn}{\end{eqnarray}}

\newcommand{\beq}{\begin{equation}}
\newcommand{\eeq}{\end{equation}}

\newcommand{\CN}{\ensuremath{\mathcal{N}}}
\newcommand{\tn}{\ensuremath{n}}
\newcommand{\ta}{\ensuremath{a}}
\newcommand{\tb}{\ensuremath{b}}

\newcommand{\tx}{\ensuremath{x}}
\newcommand{\ty}{\ensuremath{y}}
\newcommand{\ti}{\ensuremath{I}}
\newcommand{\tj}{\ensuremath{J}}
\newcommand{\tk}{\ensuremath{K}}

\def\da{{\dot \alpha}}
\def\db{{\dot \beta}}

\def\a{\alpha}
\def\b{\beta}

\newcommand{\bea}{\begin{eqnarray}}
\newcommand{\eea}{\end{eqnarray}}
\newcommand{\ee}{\end{equation}}
\newcommand{\be}{\begin{equation}}
\def\ha{{\hat \alpha}}
\def\hb{{\hat \beta}}
\def\hg{{\hat \gamma}}
\def\hd{{\hat \delta}}

\title{Unified non-metric $(1,0)$ tensor-Einstein supergravity theories  and $(4,0)$ supergravity in six dimensions}
\author[a,b]{Murat G\"{u}naydin}
\affiliation[a]{Stanford Institute for Theoretical Physics and Department of Physics, Stanford University, Stanford,
CA 94305, USA }
\affiliation[b]{Institute for Gravitation and the Cosmos \\
Pennsylvania State University\\
University Park, PA 16802, USA}
\emailAdd{mgunaydin@psu.edu}
\abstract{
The ultrashort unitary $(4,0)$ supermultiplet of $6d$ superconformal algebra $OSp(8^*|8)$  reduces to the CPT-self conjugate supermultiplet of $4d$ superconformal algebra $SU(2,2|8)$ that represents the fields of maximal $N=8$ supergravity. The graviton in the $(4,0)$ multiplet is  described by a mixed tensor gauge field which can not be identified with the standard metric in $6d$. Furthermore the $(4,0)$ supermultiplet can be obtained as a double copy of $(2,0)$ conformal supermultiplet whose interacting theories are non-Lagrangian.  It had been suggested that an interacting non-metric $(4,0)$ supergravity theory might describe the strongly coupled phase of $5d$ maximal supergravity.  In this paper we study the implications of the existence of an interacting non-metric $(4,0)$ supergravity in $6d$. The $(4,0)$ theory can be truncated to non-metric $(1,0)$ supergravity coupled to 5,8 and 14 self-dual tensor multiplets that reduce to three of the unified magical  supergravity theories in $d=5$.
This implies that the three infinite families of unified $N=2$ , $5d$ Maxwell-Einstein supergravity theories (MESGTs) plus two sporadic ones must have uplifts to unified non-metric $(1,0)$ tensor Einstein supergravity theories (TESGT)  in $d=6$. These theories have non-compact global symmetry groups under which all the self-dual tensor fields including the gravitensor transform irreducibly.
 Four of these theories are uplifts of the  magical supergravity theories  whose scalar manifolds are symmetric spaces. The scalar manifolds of the other unified theories are not homogeneous spaces.
 We also discuss the exceptional field theoretic formulations of non-metric unified $(1,0)$ tensor-Einstein supergravity theories and conclude with speculations concerning the existence of  higher dimensional  non-metric supergravity theories that reduce to the $(4,0)$  theory in $d=6$ .  }

\arxivnumber{2009.01374}
\begin{document}
\maketitle




\renewcommand{\theequation}{\arabic{section}.\arabic{equation}}
\section{Introduction}
Conformal supergravity theories with local Lagrangians based on the conformal superalgebras $SU(2,2|N)$ have long been known to exist for $N\leq 4$. It was generally believed that one could not go beyond $N=4$ without having higher spins ($>2)$.
In \cite{Gunaydin:1984vz} it was shown that the fields of maximal $N=8$ supergravity of Cremmer and  Julia \cite{Cremmer:1979up}  can be fitted into an ultra short CPT-self-conjugate unitary supermultiplet of $N=8$ superconformal algebra $SU(2,2|8)$ referred to as  the doubleton supermultiplet. The corresponding ultra short supermultiplet of $SU(2,2|4)$ is the Yang-Mills supermultiplet in $d=4$ \cite{Gunaydin:1984fk}. The $N=4$ Yang-Mills theory  of doubleton supermultiplets of $SU(2,2|4)$ is conformally invariant both classically and quantum mechanically.
This led the authors of  \cite{Gunaydin:1984vz} to pose the  question whether  a conformal supergravity theory based on the doubleton supermultiplet of $SU(2,2|8)$  exists which is closely related to
the maximal $N=8$  supergravity theory of Cremmer, Julia and Scherk. Since the latter theory is not conformally invariant  any superconformal theory based on the doubleton supermultiplet of $SU(2,2|8)$ must be unconventional or exotic.

The superalgebra $SU(2,2|8)$ was used to classify the counterterms in maximal  supergravity in \cite{Beisert:2010jx}. Furthermore, it is known that amplitudes of maximal  supergravity are $SU(8)$ covariant even though the Lagrangian does not have $SU(8)$ symmetry. This and above mentioned results provided part of the motivation for the work of Chiodaroli, Roiban and the current  author \cite{Chiodaroli:2011pp} who studied the connection between maximal supergravity  and superconformal symmetry in all dimensions that admit simple superconformal algebras as classified by Nahm \cite{Nahm:1977tg}. They showed that the six dimensional counterpart of the doubleton supermultiplet of $SU(2,2|8)$  is the $(4,0)$ supermultiplet of the superconformal algebra $OSp(8^*|8)$ with the even subalgebra $SO^*(8) \oplus USp(8)$, where $USp(8)$ is the R-symmetry group, which reduces to the CPT-self-conjugate doubleton supermultiplet of $SU(2,2|8)$ under dimensional reduction. They also showed that the $(4,0)$ theory can be obtained as a double copy of the $(2,0)$ theory based on the CPT-self-conjugate doubleton supermultiplet of $OSp(8^*|4)$\footnote{Double copy construction of the $(4,0)$ theory in terms of the $(2,0)$ theory was later reformulated by  showing  that it can be obtained as the square of  $(2,0)$ Abelian theory by using an involutive field-theoretic product at the linearized level\cite{Borsten:2017jpt}.}. The $(2,0)$  supermultiplet first appeared in the work of  \cite{Gunaydin:1984wc} who constructed the entire Kaluza-Klein spectrum of 11-dimensional supergravity over $AdS_7\times S^4$ by simple tensoring of the $(2,0)$ doubleton supermultiplet.   In the mid 1990s interacting  $(2,0)$ supersymmetric theories in $6d$ were investigated within the framework of M/Superstring theory \cite{Witten:1995zh,Seiberg:1996qx,Seiberg:1996vs, Seiberg:1997zk}.
In particular, Seiberg pointed out the existence of  four infinite series of new quantum theories with super-Poincare symmetry in
six dimensions, which are not local quantum field theories\cite{Seiberg:1997zk}.
 Later an interacting $(2,0)$ superconformal theory was proposed by  Maldacena as being dual to  M-theory on  $AdS_7\times S^4$ \cite{Maldacena:1997re}.

 The $(4,0)$ supermultiplet was studied earlier by Hull using the formalism of  double  gravitons whose equivalence to the $(4,0)$ supermultiplet obtained using the twistorial oscillators was shown in \cite{Chiodaroli:2011pp}.  Hull  argued that an interacting $(4,0)$ theory in $d=6$ might arise as the effective theory of the strongly coupled phase of five dimensional maximal supergravity when one of the dimensions decompactifies\cite{Hull:2000zn,Hull:2000rr,Hull:2001iu}.  On the other hand the interacting $(2,0)$ theory in six dimensions is believed to describe the strong coupling limit of $5d$ maximal super Yang-Mills theory. Since the maximal supergravity  can be obtained as double copy of maximal super Yang-Mills theory  in $5d$ these two proposals are consistent with the result that  $(4,0)$ theory can also be obtained as double copy of $(2,0)$ theory in $6d$  \cite{Chiodaroli:2011pp,Borsten:2017jpt}. More recently, the  action for the free $(4,0)$ theory was written down by Henneaux, Lekeu and Leonard using the formalism of prepotentials in \cite{Henneaux:2017xsb} based on their earlier work on  (2,2) mixed chiral tensor describing the graviton \cite{Henneaux:2016opm}.
The most unorthodox property of the $(4,0)$ doubleton supermultiplet of $OSp(8^*|8)$ is the fact that the field strength of the graviton does not arise from a metric and hence the corresponding theory in $6d$  is sometimes referred to as non-metric , exotic or generalized supergravity. However under dimensional reduction it reduces to the standard maximal supergravity in five and four dimensions.

Independently of the work on maximal supergravity, five dimensional  $N=2$ supergravity theories coupled to vector multiplets (MESGT) were constructed in \cite{Gunaydin:1983rk,Gunaydin:1983bi,Gunaydin:1986fg} and their gaugings were studied in \cite{Gunaydin:1984ak,Gunaydin:1984nt,Gunaydin:1984pf,Gunaydin:1999zx,Gunaydin:2000xk,Gunaydin:2000ph}. Among these MESGTs four are very special in the sense that they are unified theories with symmetric scalar manifolds $G/H$ such that $G$ is a symmetry of the Lagrangian. They were called magical supergravity theories since their symmetry groups in five , four and three dimensions
coincide with the symmetry groups of the famous Magic Square of Freudenthal, Rosenfeld and Tits \cite{Gunaydin:1983rk}. Later it was shown that there exist three infinite families of unified MESGTs and two isolated ones \cite{Gunaydin:2003yx}. Three of the magical supergravities belong to the three infinite families.  The scalar manifolds of unified MESGTs outside the magical ones are not homogeneous.  One infinite family of unified MESGTs can be gauged to obtain an infinite family of unified Yang-Mills Einstein supergravity theories in $d=5$ with the gauge group $SU(N,1)$ \cite{Gunaydin:2003yx}.

In this paper we study some of the implications of the existence of an interacting $6d$, non-metric $(4,0)$ supergravity theory. We show that the $(4,0)$ supergravity can be truncated consistently to non-metric $(1,0)$ supergravity coupled to 14, 8 and 5 self-dual tensor multiplets such that the resulting non-metric tensor-Einstein supergravity theories are unified theories  in the sense that all the tensor fields including the gravitensor transform irreducibly under a simple global symmetry group. This in turn implies that all the three infinite families of unified $5d$ MESGTs as well as the two sporadic ones must also admit uplifts  to non-metric unified (1,0) tensor-Einstein supergravity theories in $d=6$.
We conclude with some speculations about  the possible extensions of the  non-metric supergravity theories to higher dimensions with exotic spacetime signatures and the role of generalized superconformal algebras of these spacetimes.

The plan of the paper is as follows. In section 2 we review the $5d$ , $N=2$ Maxwell-Einstein supergravity theories and their gaugings. Section 3 reviews  the  truncations of $5d$, $N=8$ supergravity to three of the magical supergravity and the symmetries of octonionic magical supergravity which can not be obtained from maximal supergravity. Section 4 reviews the uplifts of magical supergravity theories to six dimensions as Poincare supergravities. In subsection \ref{superfield}  we review the on-shell superfield formulation of  $(4,0)$ supermultiplet and the gauge potentials  in the "first order formalism"  following \cite{Chiodaroli:2011pp} and give  the gauge potential of the graviton field strength in the "second order formalism". In subsection \ref{ExFT_4_0 } we review the exceptional field theoretic
formulation of linearized $(4,0)$ supergravity following the recent work of \cite{Bertrand:2020nob}. Subsection \ref{conformal} is devoted to the question whether interacting conformal supergravity theories with $SU(2,2|8)$ symmetry in $d=4$ and $OSp(8^*|8)$ symmetry in $d=6$ exist.
In section 6 we give  the truncations of $(4,0)$ supergravity to non-metric $(3,0)$ supergravity, to non-metric $(2,0)$ supergravity coupled to $(2,0)$ tensor multiplets  and to non-metric $(1,0)$ supergravity coupled to $(1,0)$ tensor multiplets. In section 7 we discuss the metric and non-metric $(1,0)$ magical supergravity theories. Section 8 is devoted general unified  non-metric $(1,0)$ tensor-Einstein supergravity theories in six dimensions and their exceptional field theoretic formulation.  In section 9 we speculate about the  possible extensions of non-metric supergravity theories to higher dimensions with non-standard space-time signatures. Appendix A reproduces the CPT-self-conjugate doubleton supermultiplet of $SU(2,2|8)$ \cite{Gunaydin:1984vz}.

\section{Review of $5D$, $\mathcal{N}=2$ Maxwell-Einstein supergravity theories \label{unifiedMESGT} and their gaugings }
\setcounter{equation}{0}

 $\mathcal{N}=2$ MESGTs in five dimensions describes
the coupling of  $\mathcal{N}=2$
 supergravity to an arbitrary number, $\tn$, of vector
multiplets.
The  supergravity multiplet consists of the
f\"{u}nfbein $e_{\mu}^{m}$, two gravitini $\Psi_{\mu}^{i}$
($i=1,2$) and one vector field $A_{\mu}$ (the "bare graviphoton").
On the other hand a $\mathcal{N} =2$ vector multiplet consists of  a
vector field $A_{\mu}$, two symplectic Majorana spinor fields $\lambda^{i}$ and
one real scalar field $\varphi$. The fermions in these theories  transform as doublets under the R-symmetry group $USp(2)_{R}\cong
SU(2)_{R}$ while all the bosonic  fields are $SU(2)_{R}$ singlets.

Hence the  fields
of an $\mathcal{N}=2$ MESGT can be labelled as
\begin{equation}
\{ e_{\mu}^{m}, \Psi_{\mu}^{i}, A_{\mu}^{\ti}, \lambda^{i\ta}, \varphi^{\tx}\}
\end{equation}
with
\begin{eqnarray*}
\ti&=& 0,1,\ldots, \tn\\
\ta&=& 1,\ldots, \tn\\
\tx&=& 1,\ldots, \tn.
\end{eqnarray*}
where we labelled the bare graviphoton as $A_{\mu}^{0}$.
The indices $\ta, \tb, \ldots$ and $\tx, \ty,
\ldots$ correspond to  the
flat and curved indices on  the
 scalar manifold, $\mathcal{M}$, respectively.

The bosonic part of the Lagrangian is
given by\cite{Gunaydin:1983bi}
\begin{eqnarray}\label{Lagrange}
e^{-1}\mathcal{L}_{\rm bosonic}&=& -\frac{1}{2}R
-\frac{1}{4}{\stackrel{\circ}{a}}_{\ti\tj}F_{\mu\nu}^{\ti}
F^{\tj\mu\nu}-\frac{1}{2}g_{\tx\ty}(\partial_{\mu}\varphi^{\tx})
(\partial^{\mu}
\varphi^{\ty})+\nonumber \\ &&+
 \frac{e^{-1}}{6\sqrt{6}}C_{\ti\tj\tk}\varepsilon^{\mu\nu\rho\sigma\lambda}
 F_{\mu\nu}^{\ti}F_{\rho\sigma}^{\tj}A_{\lambda}^{\tk},
\end{eqnarray}
where  $e$ is the determinant of  the f\"{u}nfbein
, $R$ is the scalar curvature and
$F_{\mu\nu}^{\ti}$ are the field strengths of Abelian vector
fields $A_{\mu}^{\ti}$.

The completely symmetric
tensor $C_{\ti\tj\tk}$, with lower indices is constant and determines the corresponding  $\mathcal{N}=2$
MESGT uniquely\cite{Gunaydin:1983bi}. The global symmetries of the Lagrangian are the same as symmetries of $C_{\ti\tj\tk}$. The $n$ dimensional scalar manifold can be identified with a hypersurface   in an $(n+1)$ dimensional ambient space with the metric
\begin{equation}\label{aij}
a_{\ti\tj}(h):=-\frac{1}{3}\frac{\partial}{\partial h^{\ti}}
\frac{\partial}{\partial h^{\tj}} \ln \mathcal{V}(h)\ .
\end{equation}
where
\begin{equation}
\mathcal{V}(h):=C_{\ti\tj\tk}h^{\ti}h^{\tj}h^{\tk}\ .
\end{equation}
with real variables $h^{\ti}$ $(\ti=0,1,\ldots,\tn)$ representing the coordinates of the ambient space.

The  scalar manifold  $\mathcal{M}$ is simply  the hypersurface $
{\cal V} (h)=1 $ and
the metric , ${\stackrel{\circ}{a}}_{\ti\tj}(\varphi)$,  of the kinetic energy term of vector fields  is simply the  restriction
$a_{\ti\tj}$ to $\mathcal{M}$:
\[
{\stackrel{\circ}{a}}_{\ti\tj}(\varphi)=a_{\ti\tj}|_{{\cal V}=1} \ .
\]

The physical requirements of unitarity  and positivity of the MESGT restrict the possible C-tensors.
The most general  $C_{\ti\tj\tk}$ that
satisfy these constraints can be brought to the form
\begin{equation}\label{canbasis}
C_{000}=1,\quad C_{0ij}=-\frac{1}{2}\delta_{ij},\quad  C_{00i}=0,
\end{equation}
where  $C_{ijk}$
 ($i,j,k=1,2,\ldots , \tn$) are completely arbitrary.
This is referred to as  the canonical basis. Arbitrariness  of  $C_{ijk}$ implies that
 for a given
number $\tn$ of vector multiplets, there exist, in general,  MESGTs with different
scalar manifolds and different global symmetries.


\subsection{ Unified Maxwell-Einstein Supergravity Theories \label{unified} }
Unified
 Maxwell-Einstein supergravity
theories in $d=5 $ are those theories with a simple global symmetry group under which
\emph{all}
the vector fields $A_{\mu}^{\ti}$, including the graviphoton,  form a single irreducible representation. With a combination of supersymmetry and global  noncompact symmetry group any  field can be transformed into any other field
within this class of theories.

Among MESGTs whose scalar manifolds are homogeneous spaces only four are unified theories.
They are defined by the four simple Euclidean Jordan algebras $J_3^{\mathbb{A}}$ of degree three defined by $3 \times 3$ Hermitian matrices over the four division algebras $\mathbb{A}$, namely the real numbers $\mathbb{R}$, complex numbers $\mathbb{C}$, quaternions $\mathbb{H}$ and octonions $\mathbb{O}$. The cubic norm defined by the C-tensor in these theories is identified with the cubic norm of the underlying Jordan algebra. They are referred to as magical supergravity theories because of the deep connection between their geometries and the geometries associated with
the ``magic square'' of Freudenthal, Rosenfeld and Tits \cite{MR0146231,MR0170974,MR0077076}.

In $N=2$ MESGTs defined by Euclidean Jordan algebras , $J$ , of degree three the scalar manifold is a symmetric space of the form
\eq
\mathcal{M}(J)= Str_0(J) / Aut(J) \en
where $Str_0(J)$ and $Aut(J)$ are the reduced structure and automorphism group of $J$, respectively.\footnote{Reduced structure group is the invariance group of the norm form of underlying Jordan algebra.}
Below we list the corresponding scalar manifolds:
\begin{eqnarray}
\mathcal{M}(J_3^{\mathbb{R}})&=& SL(3,\mathbb{R})/
SO(3)\qquad
(\tn=5)\nonumber\cr
\mathcal{M}(J_3^{\mathbb{C}})&=& SL(3,\mathbb{C})/
SU(3)\qquad
(\tn=8)\nonumber\cr
\mathcal{M}(J_3^{\mathbb{H}})&=& SU^{*}(6)/
USp(6)\qquad
(\tn=14)\nonumber\cr
\mathcal{M}(J_3^{\mathbb{O}})    &=& E_{6(-26)}/
F_{4}\qquad \qquad
(\tn=26)\nonumber
\end{eqnarray}

We should note that for MESGTs defined by Euclidean Jordan algebras of degree three such as the magical theories the C-tensor is an invariant tensor of the isometry group $Str_0(J)$ of the scalar manifold and we have
\eq
C_{IJK}=C^{IJK}
\en
where the indices $I,J,K,..$ are raised by the inverse ${\stackrel{\circ}{a}}^{\ti\tj}(\varphi)$ of the metric of kinetic energy term of vector fields.


In addition to four unified MESGTs defined by four simple Euclidean Jordan algebras of degree three there exist three infinite families of unified theories whose scalar manifolds are not homogeneous as was shown in \cite{Gunaydin:2003yx}. These three infinite families are defined by Lorentzian  Jordan algebras of arbitrary degree.

Now  $(n\times n)$ Hermitian matrices over
various division algebras form Euclidean Jordan algebras  with the symmetric Jordan product defined as 1/2 the anticommutator. Their automorphism groups  are compact groups.
Non-compact analogs of these algebras, denoted as
$J_{(q,n-q)}^{\mathbb{A}} $,  are generated by  matrices over
various  division
algebras , $\mathbb{A}= \mathbb{R},\mathbb{C}, \mathbb{H} $
for $n \geq 3$ and over  $ \mathbb{O}$ for $n\leq 3$,  \footnote{The Hermitian
$(n\times n)$ matrices
over the octonions do not form Jordan algebras for $n>3$.}
 that are Hermitian with respect to a non-Euclidean
``metric'' $\eta $ with signature $(q,n-q)$:
\begin{equation}\label{eta}
(\eta X)^\dag = \eta X  \hspace{1cm} \forall \, \, X\in J_{(q,n-q)}^{\mathbb{A}}
\ .
\end{equation}

It was shown in \cite{Gunaydin:2003yx} that  the structure constants (d-symbols) of traceless elements $T_{\ti}$ of noncompact Jordan algebras $J_{(1,N)}^{\mathbb{A}}$  with Lorentzian metric $\eta$ of signature $(1,N)$  defined as
\begin{equation}\label{dsymbols}
d_{\ti\tj\tk}\equiv  \frac{1}{2}
\textrm{tr}(T_{\ti}\{ T_{\tj}, T_{\tk}\})=
\textrm{tr} (T_{\ti}\circ (T_{\tj}\circ
T_{\tk}))
\end{equation}
satisfy the unitarity and positivity  requirements and can be identified
 with the C-tensor of
 a MESGT:
 \eq C_{\ti\tj\tk}=d_{\ti\tj\tk} \en
The resulting MESGTs are all unified (for $N
\geq 2 $ ) since all the vector fields including the graviphoton transform in a single irrep of the simple automorphism groups of  the underlying Jordan algebras
$\textrm{Aut}(J_{(1,N)}^{\mathbb{A}})$ which are also the symmetry groups of their Lagrangians.


 \begin{table}[ht]

\label{LorentzianJAs}
\begin{center}
\begin{displaymath}
\begin{array}{|c|c|c|c|c|}
\hline
~&~&~&~&~\\
J & D&  \textrm{Aut}(J)
& \textrm{No. of vector fields} & \textrm{No.  of scalars}  \\
\hline
~&~&~&~&~\\
J_{(1,N)}^{\mathbb R}& \frac{1}{2} (N+1)(N+2) & SO(N,1) &
\frac{1}{2}N(N+3) &\frac{1}{2}N(N+3)-1 \\
~&~&~&~&~\\
J_{(1,N)}^{\mathbb C} & (N+1)^2 & SU(N,1) & N(N+2) & N(N+2)-1 \\
~&~&~&~&~\\
J_{(1,N)}^{\mathbb H} &(N+1)(2N+1) & USp(2N,2) & N(2N+3) & N(2N+3)-1 \\
~&~&~&~&~\\
J_{(1,2)}^{\mathbb O} & 27& F_{4(-20)} & 26 & 25 \\
~&~&~&~&~ \\
\hline
\end{array}
\end{displaymath}
\end{center}

\caption{Simple Lorentzian Jordan algebras
$J_{(1,N)}^{\mathbb{A}}$. Second and third columns list
  their  dimensions $D$ and
automorphism groups $\textrm{Aut}
(J_{(1,N)}^{\mathbb{A}})$. Third and fourth columns list
the number of vector fields $(D-1)$
and the number of scalars $(D-2)$ in the $N=2$ MESGTs defined by them.}
\end{table}
In Table \ref{LorentzianJAs}  we list
 all the simple Lorentzian Jordan algebras of
type $J_{(1,N)}^{\mathbb{A}} $,
their  automorphism  groups and the number of vector and scalar fields in the MESGTs defined by them.

Remarkably the structure constants of the Lorentzian Jordan algebras of degree four
$J_{(1,3)}^{\mathbb R}$, $J_{(1,3)}^{\mathbb C}$ and $J_{(1,3)}^{\mathbb H}$
coincide with the C-tensors of  the magical MESGTs defined by  the Euclidean Jordan algebras
$J_{3}^{\mathbb C}$, $J_{3}^{\mathbb H}$ and $J_{3}^{\mathbb O}$, respectively\cite{Gunaydin:2003yx}.
Hence the magical
MESGTs based
on Euclidean Jordan algebras $J_{3}^{\mathbb C}$, $J_{3}^{\mathbb H}$ and $J_{3}^{\mathbb O}$
\cite{Gunaydin:1983rk}
are \emph{equivalent}
to the MESGTs defined by
the Minkowskian algebras
$J_{(1,3)}^{\mathbb R}$, $J_{(1,3)}^{\mathbb C}$ and $J_{(1,3)}^{\mathbb H}$,
respectively.
As a consequence the global symmetries of these theories get extended to the reduced structure groups $SL(3,\mathbb{C}), SU^*(6)$ and $E_{6(-26)}$  of the Euclidean Jordan algebras $J_{3}^{\mathbb C}$, $J_{3}^{\mathbb H}$ and $J_{3}^{\mathbb O}$, respectively. They   have the automorphism groups $SO(3,1), SU(3,1)$ and $USp(6,2)$ of $J_{(1,3)}^{\mathbb R}$, $J_{(1,3)}^{\mathbb C}$ and $J_{(1,3)}^{\mathbb H}$ as subgroups,
respectively. Furthermore their scalar manifolds are symmetric spaces as reviewed above, while the scalar manifolds of all the other MESGTs defined by Lorentzian Jordan algebras are not even homogeneous. The smallest magical  MESGT defined by the Euclidean Jordan algebra $J_{3}^{\mathbb R}$, does not belong to the three infinite families. In addition the octonionic Lorentzian Jordan algebra $J_{(1,2)}^{\mathbb O}$ defines a unified MESGT which does not belong to an infinite family. It has the global symmetry group $F_{4(-20)}$ with the maximal compact subgroup $SO(9)$.

\subsection{Unified $\CN=2$  Yang-Mills-Einstein    supergravity theories in five dimensions}


A unified $N=2$ Yang-Mills Einstein supergravity (YMESGT) theory is defined as a theory in which all the vector fields including the graviphoton transform in the adjoint representation of a simple non-Abelian subgroup of the global symmetry group that is gauged. Turning off the gauge coupling constant yields  a  unified MESGT under whose global symmetry group all the vectors transform irreducibly.

In \cite{Gunaydin:2003yx} the {\it complete} list of unified $N=2$ YMESGTs in $d=5$ was given. They are obtained by gauging the global $SU(N,1)$  symmetry groups of unified MESGTs defined by complex Lorentzian Jordan algebras $J_{(1,N)}^{\mathbb{C}}$ under which all the vector fields transform in the adjoint representation of $SU(N,1)$. As stated above the MESGT defined by $J_{(1,3)}^{\mathbb{C}}$ is equivalent to the MESGT defined by the Euclidean Jordan algebra $J_3^{\mathbb{H}}$ whose global symmetry is $SU^*(6)$. Gauging the $SU(3,1)=SO^*(6)$ subgroup of $SU^*(6)$ leads to the unique unified $5d$ YMESGT whose scalar manifold is a symmetric space \cite{Gunaydin:1984nt}. Again in \cite{Gunaydin:1984nt} it was shown that the dimensionless
 ratio $\frac{g^3}{\kappa}$  involving the
 non-Abelian gauge
coupling constant $g$ and the gravitational constant $\kappa$
must be quantized in the quantum theory by invariance under large gauge transformations.
The same argument extends to all unified YMESGTs since
\[ \Pi_5 (SU(N,1))=\Pi_5 (U(N))=\Pi_5(SU(N)) = \mathbb{Z} \ ,  \]
where $\Pi_5$ stands for the fifth homotopy group.

Pure YMESGTs in $d=5$ without tensor or hypermultiplets
do  not have a scalar potential.
By expanding around the base point
\begin{equation}
T_{0}=\left(
\begin{array}{cc}
a & 0 \\
0 & -\frac{a}{N}\mathbf{1}_{(N)}%
\end{array}
\right) ,
\end{equation}
where $a$ is some real number
fixed by the condition  $d_{000}=1$, one can show that the non-compact gauge fields transforming in $ N \oplus \bar{N}$  of $U(N)$ become massive by eating scalar fields and around this ground state $U(1)\times SU(N)$ remains unbroken with the $U(1)$ gauge field corresponding to the graviphoton. Spin 1/2 fields transforming in the symplectic $N \oplus \bar{N}$  also become massive and together with massive gauge fields form short BPS multiplets, with the central charge generated by the $U(1)$ factor.
\subsection{ ${\cal N} =2 $  Yang-Mills-Einstein
  Supergravity Theories coupled to tensor fields}
  \setcounter{equation}{0}
  Unified YMESGTs  are obtained by gauging the $SU(N,1)$ global symmetry groups of  unified MESGTs defined by the Jordan algebras $J_{(1,N)}^{\mathbb{C}}$. Since the Jordan algebras $J_{(1,N)}^{\mathbb{C}}$ are subalgebras of the quaternionic Jordan algebras$J_{(1,N)}^{\mathbb{H}}$  one can also gauge the $SU(N,1)$ subgroups of the global symmetry groups $USp(2N,2)$ of the MESGTs defined by  $J_{(1,N)}^{\mathbb{H}}$.
Under the automorphism group $USp(2N,2)$ of
$J_{(1,N)}^{\mathbb{H}}$,
the vector fields  transform in the $(2N^2+3N) $ dimensional
anti-symmetric symplectic traceless tensor representation. They  decompose as
\[ [(N+1)^2 -1 ] \oplus \frac{N(N+1)}{2} \oplus
\overline{\frac{N(N+1)}{2}} \] under the  $SU(N,1)$ subgroup of $USp(2N,2)$  for $N \geq 2$.
Therefore in gauging the $SU(N,1)$ subgroup
the  $N(N+1)$ non-adjoint vector fields must be
dualized to massive tensor fields satisfying odd dimensional self-duality conditions\cite{Gunaydin:1999zx}.

As for the family of unified MESGTs defined by the real Jordan algebras  $J_{(1,N)}^{\mathbb{R}}$,
 the vector fields transform in the symmetric
tensor representation of  $SO(N,1)$.
For even $N=2n$  with $N>3$ one
can gauge the $U(n)$ subgroup of $SO(2n,1)$ by dualizing the non-adjoint vector fields transforming in the reducible symplectic representation
\[ \frac{n(n+1)}{2} \oplus \overline{\frac{n(n+1)}{2}} \]
of $U(n)$ to tensor fields. For odd $N=2n+1$ ($N>3$) in gauging the $U(n)$ subgroup of
$SO(2n+1,1)$ the remaining vector fields in the reducible representation
\[ ( n \oplus \bar{n} ) \oplus (\frac{n(n+1)}{2} \oplus
\overline{\frac{n(n+1)}{2}}) \oplus ( 1 \oplus
\bar{1})   \] of $U(n)$ must be dualized to tensor fields.

In the  MESGT defined by the octonionic Jordan algebra
$J_{(2,1)}^{\mathbb{O}}$ with the global symmetry group $F_{4(-20)}$ one can gauge the $SU(2,1)$ subgroup with the
remaining vector fields transforming in the reducible representation
\[ (3 \oplus \bar{3} ) \oplus (3 \oplus \bar{3} ) \oplus
(3 \oplus \bar{3} ) \ .  \]
of $SU(2,1)$ dualized to tensor  fields.

\section{Magical supergravity theories and maximal supergravity}
\setcounter{equation}{0}

The magical Maxwell-Einstein supergravity theories defined by the real, complex and quaternionic Jordan algebras $J_3^{\mathbb{A}}$ ( $\mathbb{A}= \mathbb{R}, \mathbb{C},\mathbb{H}$ )  can all be obtained by a  consistent truncation of the maximal supergravity in $d=5,4$ and $3$ dimensions\cite{Gunaydin:1983rk}. The same is true  for their 6 dimensional uplifts as Poincare supergravities\cite{Gunaydin:2010fi}. The exceptional supergravity defined by the exceptional Jordan algebra  $J_3^{\mathbb{A}}$ on the other hand can not be obtained by a truncation of maximal supergravity. In five dimensions the U-duality group of maximnal supergravity is  $E_{6(6)}$ and that of exceptional supergravity is $E_{6(-26)}$. They can both be truncated to the $N=2$ MESGT defined by the quaternionic  Jordan algebra with the U-duality group $SU^*(6)$. Maximal supergravity can be gauged in $d=5$ with the gauge group $SU(3,1)$ and 12 tensor fields which admits an $N=2$ supersymmetric vacuum with vanishing cosmological constant \cite{Gunaydin:1985tb}. Similarly the exceptional supergravity theory  can be gauged with the gauge group $SU(3,1)$ and 12  tensor fields. The common sector of these two gauged supergravity theories is the unique unified $N=2$ YMESGT with the gauge group $SU(3,1)$ and whose scalar manifold is $SU^*(6)/USp(6)$.

Both real forms $E_{6(6)}$ and $E_{6(-26)}$ have $SU^*(6)\times SU(2)$ as subgroups under which they decompose as
\bea
27 = (15,1) \oplus (\bar{6},2) \\
78 = (35,1) \oplus (1,3) \oplus (20,2)
\eea
Under the maximal compact subgroup $USp(6)$ the above representations of  $SU^*(6)$ decompose as
\bea
6=6 \\
15= 14 \oplus 1 \\
20 = 6 \oplus 14' \\
35= 21 \oplus 14
\eea

Under the maximal compact subgroup $USp(8)$ of $E_{6(6)}$ we have the decompositions
\bea
27=27 \\
78 =36 \oplus 42
\eea
which further decompose under the $USp(6)\times USp(2)$ as
\bea
27 = (15,1) \oplus (6,2) \\
36 = (21,1) \oplus (1,3) \oplus (6,2) \\
42= (14,1) \oplus (14',2)
\eea
On the other hand maximal compact  subgroup of $E_{6(-26)}$  is  $F_4$ under which we have the decompositions
\bea
27=26\oplus 1 \\
78 = 52 \oplus 26
\eea
Under the $USp(6)\times USp(2)$ subgroup the above representations of $F_4$ decompose as
\bea
26 = (14,1) \oplus (6,2) \\
52 = (21,1) \oplus (1,3) \oplus (14',2)
\eea

The above decompositions show that restricting to the $USp(2)$ invariant subsector the spectra coincide with that of quaternionic magical theory defined by $J_3^{\mathbb{H}}$.  The global symmetry group $SU^*(6)$ of the quaternionic magical theory has the subgroup $SL(3,\mathbb{C})\times SO(2)$ where $ SL(3,\mathbb{C}) $ is the global symmetry group of the complex magical MESGT
\be
SU^*(6) \supset SL(3,\mathbb{C})\times U(1)_C
\ee
 The $U(1)_C$ invariant sector of the quaternionic theory corresponds to the consistent truncation to the complex magical theory. Similarly the global symmetry group of the complex magical theory decomposes as
 \be
SL(3,\mathbb{C}) \supset SL(3,\mathbb{R}) \times \mathbb{Z}_2
\ee
and $ \mathbb{Z}_2$ invariant subsector describes the consistent truncation to the real magical supergravity defined by $J_3^{\mathbb{R}}$.
\section{Magical   Poincare supergravity theories in six dimensions}
\setcounter{equation}{0}
Six dimensional magical supergravity theories coupled  to hypermultiplets and  their gaugings were studied in \cite{Gunaydin:2010fi}.
Magical supergravities in six dimensions describe the coupling of $(1,0)$ Poincare supergravity to  $n_T=2, 3, 5, 9$ tensor fields and  vector fields
in a definite spinor representation of $SO(n_T,1)$.  The coupling between vector fields and tensors  involve  $SO(n_T,1)$ invariant tensors
$\Gamma^I_{AB}$ that are the Dirac $\Gamma$-matrices for $n_T=2,3 $, and the Van der Waerden symbols for $n_T=5,9$. They satisfy the identities
\bea
\Gamma^{\vphantom{I}}_{I\,(AB}\Gamma^{I}_{C)D} &=& 0\;.
\label{magic}
\eea
which are simply the Fierz identities for the existence supersymmetric Yang-Mills theories in 3,4,6 and 10 dimensions. These identities follow from  the adjoint identity satisfied by the elements of simple Euclidean  Jordan algebras of degree three\cite{Sierra:1986dx}.
We reproduce their field contents  in Table \ref{tab:reality}.

\begin{table}
\begin{center}
\begin{tabular}{c||c|c|c|c}
$G_T$ & ${\cal R}_{\rm v}$ & $A_\mu^A$ & $\Gamma^I_{AB}$ & ${\cal R}_{\rm ten}$ \\[1ex] \hline\hline
&&&&\\[-2ex]
${SO}(9,1)$ &
${\bf 16}_c$
& MW & $\Gamma^I_{AB}$
& ${\bf 10}$
\\[1ex]\hline&&&&\\[-2ex]
${SO}(5,1)\times {USp}(2)$ &
${\bf (4}_c,{\bf 2)}$
& SMW,\, $A=(\alpha r)$ &
$\Gamma^I_{\alpha r,\beta s} = \Gamma^I_{\alpha\beta} \epsilon_{rs}$
& ${\bf (6,1)}$
\\[1ex]\hline&&&&\\[-2ex]
${SO}(3,1)\times {U}(1)$ &
${\bf (2,1)}_+ + {\bf (1,2)}_-$
&W,\, $A=\{\alpha,\dot{\beta}\}$
&
$\left(
\begin{array}{cc}
0 & \Gamma^I_{\alpha\dot{\beta}}\\
\bar\Gamma^I_{\dot\alpha{\beta}} & 0
\end{array}
\right)$
& ${\bf (2,2)_0}$
\\[3ex]\hline&&&&\\[-2ex]
${SO}(2,1)$  &
${\bf 2}$&
M & $\Gamma^I_{AB}$
& ${\bf 3}$
\\[1ex]\hline
\end{tabular}
\caption{
{\small Above we reproduce the field content and symmetries of magical supergravity theories in six dimensions.
The first column lists their global symmetry groups $G_T$
and the second column lists the representations ${\cal R}_{\rm v}$ of the vector fields $A_\mu^A$ under $G_T$. The reality properties of these representations are given in the third column: Majorana (M), Weyl (W), Majorana-Weyl (MW), symplectic Majorana-Weyl (SMW).
The last column lists  the representations ${\cal R}_{\rm ten}$ of the tensor fields under $G_T$. $\Gamma^I$ are the gamma matrices in the respective dimensions
 }}
\label{tab:reality}
\end{center}
\end{table}

Since the vector  fields transform in a  spinor representation which belong to a unique orbit of the isometry group of the scalar manifold one finds that six dimensional magical supergravity theories  admit a unique gauge group  which is a centrally extended Abelian nilpotent group. For the octonionic magical theory the unique gauge group is the maximal centrally extended Abelian subgroup of $F_{4(-20)}$ which is the  automorphism group of the Lorentzian octonionic Jordan algebra $J_{(2,1)}^{\mathbb{O}} $.
For the quaternionic ( complex)  magical theory the unique gauge group is the maximal centrally extended Abelian subgroup of $USp(4,2)$ ( $SU(2,1)$ ) which is the  automorphism group of the Lorentzian quaternionic ( complex)  Jordan algebra $J_{(2,1)}^{\mathbb{H}} $ ($J_{(2,1)}^{\mathbb{C}} $) .

They satisfy the inclusions

\eq
F_{4(-20)} \supset USp(4,2) \times USp(2) \supset SU(2,1) \times U(1)
\en

These results show that semisimple gaugings of the $5d$ magical supergravity theories do not admit uplifts to six dimensions as standard Lagrangian Poincare supergravities. On the other hand it is known that the $5d$ , $N=4$ super Yang-Mills theory can be obtained from $(1,1)$ Poincare supersymmetric Yang-Mills theory in six dimensions by dimensional reduction and it is generally believed that it can also be obtained from an interacting (2,0) superconformal field theory.
Similarly the $N=2$ super Yang-Mills theory in five dimensions can be obtained from $(1,0)$ supersymmetric Yang-Mills theory or from a $(1,0)$ superconformal theory of self-dual tensor multiplets in $d=6$. The standard Yang-Mills theories in $d=6$ involving vector fields are not conformally invariant.
The $(2,0)$ conformal supermultiplet decomposes as a $(1,0)$ tensor multiplet plus a conformal hypermultiplet in $d=6$
\be
(2,0) = (1,0) \,  \textrm{tensor multiplet} \oplus (1,0) \,  \textrm{hypermultiplet}
\ee
Therefore an interacting $(2,0)$ theories can be viewed as a special family of interacting $(1,0)$ tensor multiplets  coupled to hypermultiplets. Similarly the $N=4$ super Yang-Mills theories that descend from the interacting $(2,0)$ theories in $d=6$ can be viewed  as a special class of $N=2$ super Yang-Mills theories coupled to $N=2$ hypermultiplets in the adjoint representation of the gauge group.
\section{ Superconformal symmetry and non-metric  (4,0)  supergravity in six dimensions}
\subsection{On-shell superfields of $6d$ , $(4,0)$ supermultiplet of $OSp(8^*|8)$ in twistorial formulation and first versus second order formalism \label{superfield} }
The physical degrees of freedom corresponding to the fields of maximal $N=8$ supergravity in $d=4$ were shown to belong to the CPT-self-conjugate unitary representation (doubleton) of the conformal superalgebra $SU(2,2|8)$ in \cite{Gunaydin:1984vz}. Formulation of this unitary supermultiplet in terms of constrained on-shell superfields was given in \cite{Chiodaroli:2011pp} which we review in the Appendix. Even though the physical degrees of freedom form a unitary supermultiplet of the conformal superalgebra $SU(2,2|8)$ interactions in maximal supergravity break the conformal symmetry down to Poincare subgroup. Whether a conformal supergravity based on this supermultiplet exists, as contemplated in \cite{Gunaydin:1984vz}, is still an open problem as discussed below.

In six dimensions the unique superconformal algebra with 64 supersymmetry generators is $OSp(8^*|8)$ with the maximal even subalgebra $SO^*(8) \oplus USp(8)$. Explicit construction of the CPT self-conjugate doubleton supermultiplet of $OSp(8^*|8)$ in terms of twistorial oscillators was given in \cite{Chiodaroli:2011pp} and shown to reduce to the doubleton supermultiplet of $SU(2,2|8)$ under dimensional reduction.
In Table \ref{USp(8)_doubleton} we  reproduce the doubleton supermultiplet of $OSp(8^*|8)$ with
$R$-symmetry group $USp(8)$ given in \cite{Chiodaroli:2011pp}\footnote{We should note that in the manifestly unitary construction using twistorial oscillators the field strengths corresponding to physical degrees of freedom and not the corresponding gauge fields  form the unitary supermultiplets. With that caveat we will use the terms fields and field strengths interchangeably.}. This multiplet is referred to as the $(4,0)$ conformal supermultiplet in $d=6$ and  was studied earlier  by Hull using the  formalism of double gravitons who argued that an interacting theory based on this supermultiplet may describe a strongly coupled phase of $5d$ maximal supergravity when one of the dimensions decompactifies \cite{Hull:2000zn,Hull:2000rr,Hull:2001iu}.

\begin{table}[ht]
\begin{center}
\begin{tabular}{|c|c|c|}
\hline
 Field Strengths &  ${SU^{*}(4)}_{D}$ & $USp(8)$
\\ \hline

$\phi^{[ABCD]|}(x)$  & (0,0,0)       & 42     \\ \hline
$\lambda_{\hat{\alpha}}^{[ABC]|} (x) $ & (1,0,0)
& 48\\ \hline

$h_{(\hat{\alpha}\hat{\beta})}^{[AB]|}(x)$ & (2,0,0)       & 27     \\ \hline
$\psi_{(\hat{\alpha}\hat{\beta}\hat{\gamma})}^A (x) $& (3,0,0) & 8 \\ \hline
$R_{(\hat{\alpha}\hat{\beta} \hat{\gamma}\hat{\delta})}(x) $& (4,0,0) & 1 \\ \hline
\end{tabular}
\end{center}
\caption {\label{USp(8)_doubleton} \small
(4,0) doubleton supermultiplet of $OSp(8^*|8)$. First column lists the field strengths corresponding to actual physical degrees of freedom. Second column gives their Dynkin labels under the Lorentz group  $SU^*(4)$. Third column gives the  dimensions
of their $USp(8)$  representations.  The field strengths are labelled by chiral
$SU^*(4)$ spinor indices $\hat{\alpha}, \hat{\beta},\dots$ and $USp(8)$ indices $A,B,\dots$.}
\end{table}

The fields belonging to the $(4,0)$ supermultiplet  can be fitted into an on-shell
superfield satisfying an algebraic and a differential constraint \cite{Chiodaroli:2011pp}.
For this it turns out to be very convenient to represent the coordinates of the  six-dimensional
extended superspace  as anti-symmetric tensors in spinorial indices \cite{Howe:1983fr,Koller:1982cs}
\be
\big( x^{\ha \hb}= -x^{\hb \ha}, \theta^{\hat \alpha}_A \big) \qquad \ha, \hb = 1, \dots, 4 \ ;
\quad A=1, \dots, 8 \ ;
\ee
where the spinorial indices of the Lorentz group $SU^*(4)$ in $d=6$
 are labelled by hatted Greek
indices $\ha, \hb  \dots$ and the $USp(8)$  indices
by $A,B,C \dots$.
Defining the  superspace covariant derivative
\be
D^A_{\hat \alpha} = \partial^A_{\hat \alpha} +
i \Omega^{AB} \theta^\hb_B \partial_{\hat \alpha  \hb} \ ,
\ee
where $\partial^A_{\hat \alpha} \theta^\hb_B = \delta^A_B \delta^\hb_\ha$ and  $\Omega_{AB}=-\Omega_{BA}$
 one finds
\be \{ D^A_\ha, D^B_\hb \} = 2 i \Omega^{AB} \partial_{\ha \hb} \ ,
\qquad \{ D_{\ha A}, D^B_{\hb} \} = 2 i \delta^B_A \partial_{\ha \hb} \ . \ee
The symplectic metric satisfies
\be
\Omega_{AB} \Omega^{BC} = \delta^C_A \ .
\ee
and is  used to raise or lower indices,
\be
\theta^{A\ha} = \Omega^{AB} \theta^\ha_B, \qquad D_{A\ha}= \Omega_{AB} D^B_\ha \ .
\ee

The scalar  superfield  of the $(4,0)$ supermultiplet
\be
\Phi^{ABCD}\big( x^{\ha \hb}, \theta^\ha_A \big)
\ee
is completely anti-symmetric in its indices and is symplectic traceless i.e.
\be
\Phi^{ABCD} \Omega_{CD}= 0  ~~ ,
\label{algebraicconstraint}
\ee
and satisfies the
differential constraint
\be
D^A_\ha \Phi^{BCDE} + {1 \over 21} D_{\ha F} \Big\{ \Omega^{A[B} \Phi^{CDE]F}
+ {3 \over 4} \Omega^{[BC} \Phi^{DE]AF} \Big\}=0~~.
\label{diffconstraint}
 \ee

The mapping between twistorial formulation of $(4,0)$ supermultiplet and formulation in terms of vectorial indices $M,N,..=0,1,..,5$ of $SO(5,1)$ as was done by Hull was given in \cite{Chiodaroli:2011pp}.
The field strength $R_{\ha \hb \hg \hd}$ of the non-metric graviton corresponds to the $(3,3)$ tensor $R_{[MNO][PQR]}= R_{[PQR][MNO]}$ which satisfies self-duality conditions in the first as well as the last 3 indices
\be
*R=R* =R  ~~  ,
\ee
where $*$ operation is performed with the Levi-Civita tensor in six dimensional Minkowskian spacetime
\[ (*R)_{[MNO][PQR]}= \frac{1}{6} \epsilon_{MNOSTU}R^{STU}_{~~~~ PQR} ~~. \]

In \cite{Chiodaroli:2011pp} the gauge potential for the non-metric graviton field strength  $R_{\ha \hb \hg \hd}$ transforming in the $(4,0,0)_D$ representation of $SU^*(4)$ was chosen as a tensor field $C^{\ha}_{(\hb \hg \hd)}$ transforming in the $(3,0,1)_D$ representation such that the field strength involves a single derivative
\be
R_{(\ha \hb \hg \hd)}=\partial^{\vphantom{ \hat \omega \hat \lambda}}_{ \hat \lambda ( \ha}  C^{\hat \lambda }_{\hb \hg \hd)}
\ee
where
\be  C^{\hat \lambda }_{(\hb \hg \hd) } = C^{\hat \lambda }_{\hb \hg \hd } \, \quad .
 \ee

It is invariant under the gauge transformations
\be  C^{\ha}_{\hb \hg \hd} \rightarrow C^{\ha}_{\hb \hg \hd} +
\partial_{ \hat \omega ( \hb }^{\vphantom{\ha}} \chi^{\hat \omega \ha}_{\hg \hd)} \ee
where the gauge parameters $\chi^{\ha \hb}_{\hg \hd}$ satisfy
\be
\chi^{\ha \hb}_{\hg \hd} = - \chi^{\hb \ha}_{\hg \hd} = \chi^{\ha \hb}_{\hd \hg} \, \quad .
\ee

 However, since the standard  Riemann tensor involves  two derivatives of the metric one can also choose a gauge potential such that the non-metric graviton field strength  involves two derivatives of that gauge potential as was done in \cite{Hull:2000rr,Henneaux:2017xsb}.
 In terms of spinorial indices such a gauge potential  must transform  as a mixed tensor $ C^{ \ha \hb }_{ \hg \hd} $
  satisfying the conditions
  \bea
  C^{ \ha \hb }_{ \hg \hd}=C^{ \hb \ha }_{ \hg \hd}=C^{ \ha \hb }_{ \hd \hg} \\
  C^{ \ha \hb }_{ \hg \hb}=0
  \eea
 such that  the non-metric graviton field strength is given by \be
R_{(\ha \hb \hg \hd)}=\partial^{\vphantom{ \hat \omega \hat \lambda}}_{ \hat \lambda ( \ha}  \partial^{\vphantom{ \hat \omega \hat \lambda}}_{ \hat \epsilon  \hb} C^{\hat \lambda \hat \epsilon}_{\hg \hd )}  \label{secondorder}
\ee
which is invariant under the gauge transformations
\be
 C^{\ha \hb}_{\hg \hd } \longrightarrow C^{\ha \hb}_{\hg \hd } + \partial_{\hat \lambda ( \hg } \chi^{\ha \hb \hat \lambda}_{\hd )}
 \ee
 where the gauge parameters satisfy
 \bea
  \chi^{\ha \hb \hat \lambda}_{\hd }=- \chi^{\ha \hat{\lambda}  \hb}_{\hd } \\
   \chi^{\ha \hb \hg }_{\hg } =0
   \eea
 which imply that they transform in the 64 dimensional  representation of $SU^*(4)$ with Dynkin label  $(1,1,1)_D$. The formulation given in \cite{Chiodaroli:2011pp} and the formulation in terms of a gauge potential of the form  \ref{secondorder} given in \cite{Hull:2000rr,Henneaux:2017xsb} are the analogs of first and second order formalisms in ordinary supergravity \cite{Freedman:2012zz}.\footnote{ I would like to thank Marc Henneaux for posing the question about the difference between the formulations in  \cite{Chiodaroli:2011pp} and in \cite{Hull:2000rr,Henneaux:2017xsb} that triggered the investigations that led to this paper. } The underlying unitary $(4,0)$ supermultiplet  that describes the physical degrees of freedom is the same for both formulations  just as is the case for first and second order formalism of Poincar\'e supergravity.

   The gauge potential of the non-metric gravitino field strength is a traceless tensor
$\psi^{\ha}_{(\hb \hg)}$.
 Under a gauge transformation it transforms as
\be
\psi^{\ha }_{\hb \hg}  \rightarrow \psi^{\ha}_{\hb \hg} +
\partial_{\hat \omega  ( \hb } \chi^{\hat \omega \ha }_{\hg)} \ ;
\ee
with the gauge parameter $\chi^{[\ha \hb]}_\hg$ such that  $\chi^{\ha \hb}_\hb =0$.
We should note that  in contrast to standart local supersymmetry gauge parameter which involves a single spinor index, the gauge parameter $\chi^{\ha \hb}_\hg$ of  this  local gauge symmetry of the non-metric  gravitino field
transforms as a spinor vector under $SU^*(4)$.

In terms of vectorial indices the non-metric  gravitino field can be written as  $\psi_{[MN]\ha}$ such that its  field strength $\psi_{[MNO]\ha} $ satisfies the self-duality condition
\be
* \psi = \psi
\ee
Similarly the tensor field with field strength $h_{(\ha \hb)}$ is $ b^\ha_\hb $ so that
\be
 h_{(\ha \hb)}= \partial^{\vphantom{\hg}}_{\hg (\ha} b^{\hg}_{\hb)} \, , \quad b^\ha_\ha=0 \ee
 It undergoes gauge transformations with parameters $\chi^{\ha\hb}=-\chi^{\hb\ha}$.
 In terms of vectorial indices it is described by an anti-symmetric tensor field $b_{[MN]}$ whose field strength $h_{[MNP]}$ satisfies the selfduality condition
 \[ *h =h \]

\subsection{ A Review of Exceptional Field Theoretic Formulation of the linearized $(4,0)$ Supergravity \label{ExFT_4_0 } }
The fact that  the ultrashort (4,0) supermultiplet of $N=8$ , $d=6$ superconformal algebra $OSp(8|8^*)$ reduces to the ultrashort doubleton supermultiplet of $N=8$  $4d$ superconformal algebra $SU(2,2|8)$ that describes the fields of $4d$ maximal supergravity raises several interesting questions. First is the question whether there exist interacting supergravity theory or theories  based on this supermultiplet. In four dimensions interactions in  maximal Poincare  supergravity break the superconformal symmetry $SU(2,2|8)$ down to its Poincare subalgebra.
On the other hand $N=4$ super Yang-Mills theory based on the corresponding ultrashort unitary supermultiplet of $PSU(2,2|4)$ is conformally invariant even at the quantum level. Remarkably the amplitudes of maximal $N=8$  Poincare supergravity can be obtained by double copy methods of Bern, Carrasco and Johansson (BCJ)  \cite{Bern:2008qj,Bern:2010ue}  from those of $N=4$ super Yang-Mills theory that is conformally invariant \cite{Bern:2009kd}\footnote{For a review and references on the subject see \cite{Bern:2019prr}.}.

Existence of an interacting $4d$ Poincare supergravity based on the conformal ultrashort CPT self-conjugate supermultiplet of $SU(2,2|8)$  suggests that an interacting Poincare supergravity in $6d$ based on the conformal $(4,0)$ supermultiplet of $OSp(8^*|8)$ also exists in which the interactions break the superconformal symmetry $OSp(8^*|8)$ down to its Poincare subgroup. On the other hand  the existence of conformal supergravity based on the $(4,0)$ supermultiplet in six dimensions is an open question just like  the existence of an interacting $N=8$ conformal supergravity in
$d=4$ as will be discussed below.

In six dimensions the $(2,0)$ supermultiplet of $OSp(8^*|4)$ is the conformal analog of $4d$ $N=4$ Yang-Mills supermultiplet \cite{Gunaydin:1984wc}. It is generally believed that the interacting theories of $(2,0)$ supermultiplets in six dimensions are not conventional field theories and may only exist as quantum theories\cite{Witten:2007ct}. Nonetheless they reduce to conventional field theories in lower dimensions. In \cite{Chiodaroli:2011pp} it was pointed out that the $(4,0)$ supermultiplet can be obtained by tensoring the $(2,0)$ supermultiplets. This raises the possibility that the "amplitudes" or  correlation functions of an interacting superconformal $(2,0)$ theory could yield the amplitudes of an interacting non-metric  Poincare supergravity based on the $(4,0)$ supermultiplet via some generalization of double copy methods of BCJ.

Even though the supermultiplet of fields of $(4,0)$ supergravity and  their free equations were known for a long time
the action for linearized non-metric $(4,0)$ Poincare supergravity in six dimensions was first written down rather recently  in \cite{Henneaux:2017xsb}. The authors of \cite{Henneaux:2017xsb} use  the formalism of prepotentials  adapted to the self-duality properties of the fields of the $(4,0)$ supermultiplet.
They show that the resulting  action is invariant under $(4,0)$ Poincar\'e supersymmetry in $d=6$ but not manifestly. The reason for loss of manifest Poincar\'e covariance is due to the fact that to write down the action they split the $6d$ spacetime coordinates as $5+1$ with the singlet coordinate being timelike. In their
Lagrangian formulations of the bosonic self-dual tensors which they refer to as chiral two-forms \cite{Henneaux:1988gg}  as well as bosonic chiral
(2,2)-tensor corresponding to the gauge potential of non-metric  graviton \cite{Henneaux:2016opm}  involve only spatial tensors,
and their temporal components  are pure gauge\footnote{With the caveat that
equations of motion do not yield  the self-duality conditions directly but
an equivalent differential form.}. They also present  a  similar formulation for the non-metric gravitensorino (" chiral spinorial two-form ").
The resulting action of free  $(4,0)$ supergravity in terms of prepotentials is fourth order in spatial derivatives.\footnote{The authors of \cite{Henneaux:2017xsb} also present the action for the free $(3,1)$ supergravity. Unlike the $(4,0)$ supermultiplet $(3,1)$ Poincare supermultiplet does not extend to a conformal supermultiplet in six dimensions.}

$(4,0)$ supergravity as well as the $(3,1)$ supergravity were also studied within the so-called exceptional field theory (ExFT) formalism   by the authors of \cite{Bertrand:2020nob} recently. Before giving  the exceptional field theoretic formulation of these theories they first present  novel actions for the bosonic sectors of linearized  $(4,0)$ and $(3,1)$ using the $5+1$ split of six dimensional spacetime coordinates such that the singlet coordinate $y$ is space-like. These actions are two-derivative actions that reduce to the bosonic sector of  linearized maximal supergravity in five dimensions.

The  exceptional field theory formalism  is a particular extension of the double field theory  formalism and is an outgrowth of the attempts to make the hidden U-duality groups of lower dimensional supergravity theories manifest in  higher dimensions  from which they can be obtained by toroidal compactification\footnote{ For a review and references on the subject see  \cite{Hohm:2019bba}.}. To achieve this one intoduces an auxiliary "internal" spacetime with coordinates $Y^M$ motivated by the U-duality group with which to extend the standard external d-dimensional spacetime with coordinates $x^\mu$ and imposes a section constraint such that the resulting theory describes the higher dimensional supergravity theory.  For 5d maximal supergravity one introduces a 27 dimensional auxiliary internal space-time extending the 5 dimensional external spacetime and imposes  a section constraint of the form
\be
C^{IJK} \partial_J \otimes \partial_K =0
\ee
where $C^{IJK}$ is the $E_{6(6)}$ invariant symmetric tensor.  The above equation is to be interpreted as differentials acting on functions $f(x,Y), g(x,Y)$ of extended coordinates $x^\mu, Y^I$ such that
\be
C^{IJK} \partial_J \partial_K f(x,Y) =0  \quad , \quad C^{IJK} \partial_J f(x,Y) \partial_K g(x,Y) =0
\ee
To recover 11-d supergravity corresponding to the $5+6$ split of the coordinates one decomposes the 27 internal coordinates $Y^I$ with respect to $SL(6,\mathbb{R})\times GL(1) $ subgroup of $E_{6(6)}$
\be
27 = 6^{+1} +15^0 + 6^{-1} \Leftrightarrow Y^I = (y^m , y^{mn}=-y^{nm}, \bar{y}^m )
\ee
where $m,n,..=1,..,6$.
By restricting the dependence on $Y^I$ only to the 6 coordinates $y^m$ one obtains a solution to section constraints that leads to the 11 dimensional supergravity\cite{Hohm:2013vpa}.\footnote{Type IIB supergravity corresponds to the decomposition of 27 of $E_{6(6)}$ with respect to the $SL(5,\mathbb{R})\times SL(2,\mathbb{R}\times GL(1)$ subgroup given by $27 = (5,1)^{-4} + (5',2)^{-1} +(10,1)^{-2} +(1,2)^-5$ and restricting dependence on internal coordinates to  the coordinates $y^a$ in  $(5,1)$. }

To construct the linearized 6d $(4,0)$ , $(3,1)$ and standard $(2,2)$ supergravity theories  as  ExFTs in a unified manner describing the maximal $N=8$ supergravity in $5d$ the authors of \cite{Bertrand:2020nob} extend the 27 dimensional  internal space-time with  an extra singlet coordinate $Y^\bullet $ and impose the more general section constraint
\be
C^{IJK} \partial_J \otimes \partial_K - \frac{1}{\sqrt{10} } \Delta^{IJ} ( \partial_J \otimes \partial_\bullet +\partial_\bullet \otimes \partial_J ) =0 \label{gensection}
\ee
where $\Delta^{IJ}$ is a constant tensor describing the background spacetime. Setting $\partial_\bullet =0$ one has the standart section constraint of the formulation of the maximal $N=8$ supergravity as an ExFT whose solutions include the 11d sugra, type IIB supergravity and maximal $(2,2)$ Poincare supergravity theory in $6d$. On the other hand setting $\partial_I =0$ the generalized section constraint \ref{gensection}  is satisfied trivially and by identifying the extra coordinate $Y^\bullet$ with the space-like coordinate in the $(5+1)$ split of $6d$ coordinates $x^\mu$ they show that one obtains the bosonic sector of  linearized $(4,0)$ supergravity. In addition to $(2,2)$ and linearized $(4,0)$ supergravity in $6d$  the generalized section constraint admits a solution corresponding to $(3,1)$ supergravity at the linearized level as well.   They leave to future work the supersymmetric extension of the bosonic sector of $(4,0)$ supergravity. Furthermore the fact that a unified framework exists for ExFT formulations of $6d$ (2,2), (4,0) and (3,1) supergravity theories and the $(2,2)$ theory can be extended to the full nonlinear theory is interpreted by the authors of \cite{Bertrand:2020nob} as evidence that the same may be true for the $(4,0)$ and $(3,1)$ theories. If the interacting non-metric $(4,0)$ supergravity exists as a Lagrangian theory it will admit a formulation as an ExFT describing the uplift of maximal Poincare supergravity to six dimensions. However if the interacting $(4,0)$ theory is non-Lagrangian we shall assume that an appropriate generalization of the ExFT formalism exists with which to uplift the $5d$ maximal supergravity to $6d$ as a non-metric $(4,0)$ Poincare supersymmetric  theory.

\subsection{ On the question of existence of interacting $N=8$ conformal supergravity theory in $d=4$  and  $(4,0)$ conformal supergravity in  $d=6$ \label{conformal} }
The existence of an interacting conformal supergravity based on the $(4,0)$ supermultiplet in $d=6$ would suggest the existence of an interacting  conformal supergravity in $d=4$ based on the CPT self-conjugate doubleton supermultiplet of $SU(2,2|8)$ whose existence was posed as an open problem in \cite{Gunaydin:1984fk}. To  this date no such supergravity theory has been constructed.
Conformal supergravity theories were first studied in the pioneering papers of \cite{Kaku:1977pa,Ferrara:1977ij}.\footnote{ For an older review see \cite{Fradkin:1985am}.}
Recently they have been studied as massless limits of Einstein-Weyl supergravity theories\footnote{ For a  review  see \cite{Ferrara:2020zef} and the references therein. }.
 The standard conformal supergravity theories in $d=4$ based on the conformal superalgebras $SU(2,2|N)$ exist only for $N\leq 4$.
  All N =4 conformal supergravities in $d=4$  have recently been constructed in \cite{Butter:2016mtk,Butter:2019edc}.
  In addition to a massless graviton they contain a massive spin two ghost field and hence are not unitary. The constraint $N\leq 4$ arises from the fact that for $N\geq 4$ the conformal supermultiplets containing the massive spin two field must necessarily contain fields of spin greater that two. Furthermore  the vector fields associated with the gauge fields of $SU(n)$  and $U(1)$   subgroups inside  $U(n) \subset SU(2,2|n)$ have kinetic energy terms that are of opposite sign and hence are not all positive definite.  Hence if an  interacting $N=8$ superconformal theory exists that is unitary its formulation must go beyond the standard local gauging of the underlying conformal superalgebras. It may exist  purely at the quantum level without a Lagrangian formulation or its Lagrangian formulation may be non-local. In fact there are non-local formulations of conformal gravity that are both unitary and ultraviolet finite\footnote{For reviews we refer to \cite{Modesto:2017sdr,Rachwal:2018gwu} and references therein.}. Their Lagrangians  typically involve infinite number of terms that are bilinear in scalar curvature $R$, Ricci tensor $R_{\mu\nu}$ and Riemann tensor $R_{\mu\nu\rho\lambda}$ with powers  of the D'Alembertian sandwiched between them.   These results are consistent with the findings of \cite{Gording:2018not}  who studied the four derivative action of Stelle\cite{Stelle:1977ry} of the form
  \be
  S= M_{Pl}^2 \int  d^4x \sqrt{-g} \left[ -2 \Delta + R +\frac{1}{\mu^2} ( \frac{1}{3} R^2 - R^{\mu\nu} R_{\mu\nu} ) \right]
  \ee
  where $M_{Pl}$ is the Planck mass, $\Delta$ is the cosmological constant and $\mu$ indicates the mass scale. This theory has a massive spin two ghost of mass $\mu$ whose kinetic energy has opposite sign to that of the massless mode. They show that there exists a  ghost free completion of this theory which requires an infinite series of higher derivative terms. The resulting theory is classically equivalent to ghost-free bimetric theory of two symmetric tensor fields studied in \cite{Hassan:2011zd}.
  Whether these ultraviolet finite unitary nonlocal (higher derivative) theories admit supersymmetric extensions that go beyond the $N=4$ bound and result in a unitary theory whose massless physical spectrum coincides with the CPT-self-conjugate doubleton supermultiplet of $SU(2,2|8)$ is an open problem.

The question about the existence of  interacting conformal supergravity theory with $OSp(8^*|8)$ symmetry in $d=6$ is more subtle. Such a theory would necessarily involve a non-metric "graviton" and no interacting non-metric gravity theories have been constructed to date.
The metric conformal  supergravity theories in $d=6$ with superconformal symmetry $OSp(8^*|2N)$ exist only for $N\leq 2)$. Whether there exist non-local ( higher derivative) and non-metric conformal supergravity theories with $OSp(8^*|8)$ symmetry  and  are unitary is an open problem.

\section{ Truncations of $(4,0)$ supergravity}
\setcounter{equation}{0}
Assuming that there exists an interacting   $(4,0)$ supergravity that reduces to the maximal $N=8$ supergravity one can consider its
 truncations to  interacting theories with lower number of supersymmetries.
\subsection{ $(2,0)$ supersymmetric truncations}
The decomposition of the $(4,0)$ supermultiplet of $OSp(8^*|8)$ into supermultiplets of $N=4$  superconformal algebra $OSp(8^*|4)$ in $d=6$ and some of its implications were studied in \cite{Chiodaroli:2011pp}. One finds that $(4,0)$ multiplet decomposes into  one $(2,0)$ non-metric graviton supermultiplet\cite{Chiodaroli:2011pp}
\eq
R_{(\hat{\alpha}\hat{\beta} \hat{\gamma}\hat{\delta})}(x),  \quad  \psi_{(\hat{\alpha}\hat{\beta}\hat{\gamma})}^a (x),  \quad  h_{(\hat{\alpha}\hat{\beta})}^{[ab]|}(x) \oplus h_{(\hat{\alpha}\hat{\beta})}(x) , \quad \lambda_{\hat{\alpha}}^{[abc]|} (x) , \quad \phi(x)
\label{confgrav6d}\en
four $(2,0)$  non-metric gravitino supermultiplets
\eq
 \psi_{(\hat{\alpha}\hat{\beta}\hat{\gamma})} (x),  \quad  h_{(\hat{\alpha}\hat{\beta})}^a(x) , \quad \lambda_{\hat{\alpha}}^{[ab]|} (x) \oplus \lambda_{\hat{\alpha}} (x) , \quad \phi^{[abc]|}(x)
\en
and five  $(2,0)$ self-dual tensor multiplets ( doubletons )
\eq
  h_{(\hat{\alpha}\hat{\beta})}(x) , \quad \lambda_{\hat{\alpha}}^{a} (x)  , \quad \phi^{[ab]|}(x)
\en
where   $a,b,c=1,2,3,4$ are the $USp(4)$ indices.
Therefore the interacting $(4,0)$ theory must admit a consistent truncation that  describes the coupling of the non-metric $(2,0)$ graviton supermultiplet to five
$(2,0)$ doubleton supermultiplets.

The $(2,0)$ supersymmetric truncation of the $(4,0)$ conformal supergravity  reduces to $N=4$ supergravity coupled to 5 vector multiplets in $d=5$ with the global symmetry group $SO(5,5)\times SO(1,1)$ which is also the global symmetry group of the 6 dimensional theory with the moduli space
\[ SO(5,5)\times SO(1,1)/ SO(5)\times SO(5)
\]
 The tensor fields of the six dimensional theory form the  (10+1) representation of $SO(5,5) \times SO(1,1)$.  The resulting supergravity can be gauged in $d=5$ to obtain  Yang-Mills Einstein supergravity theories with various possible gauge groups, in particular $SU(2)\times U(1)$. Since it is generally believed that strongly coupled phase of  $5d$ maximal super Yang-Mills is described by  an interacting $(2,0)$ theory in $d=6$ this raises the question whether there exist deformations of  the non-linear $(4,0)$ theory that correspond  to various gaugings of maximal supergravity in $d=5$ or in $d=4$. In general not all gaugings of maximal supergravity are expected to have uplifts to higher dimensions since gaugings in general introduce potentials with anti-de Sitter  as well as de Sitter vacua.
\subsection{ $(1,0)$ supersymmetric truncations of $(4,0)$ conformal supergravity }
\setcounter{equation}{0}
The $(4,0)$ supermultiplet of $OSp(8^*|8)$ can be decomposed into $(1,0)$ supermultiplets. Splitting the $USp(8)$ indices $A,B,..$ as $A=(a,i), B=(b,j),..$ where $i,j,..=1,2$ and $a,b,..=3,4,..,8$ we find
\begin{itemize}
\item non-metric graviton supermultiplet
\[  R_{(\hat{\alpha}\hat{\beta} \hat{\gamma}\hat{\delta})}(x),  \quad  \psi_{(\hat{\alpha}\hat{\beta}\hat{\gamma})}^i (x),  \quad  h_{(\hat{\alpha}\hat{\beta})}^{[ij]}(x) \]

\item 6 non-metric gravitino ( gravitensorino) supermultiplets \[
\psi^a_{(\hat{\alpha}\hat{\beta}\hat{\gamma})} (x),  \quad  h^{ai}_{(\hat{\alpha}\hat{\beta})}(x) , \quad \lambda_{\hat{\alpha}}^{a[ij]} (x)  \]
\item
14 (1,0) self-dual tensor multiplets
\[
  h_{(\hat{\alpha}\hat{\beta})}^{[ab]|} (x) , \quad \lambda_{\hat{\alpha}}^{[ab]|i} (x)  , \quad \phi^{[ab]|ij}(x)
\] transforming in the antisymmetric symplectic traceless representation of $USp(6)$.
\item
14 (1,0) hypermultiplets
\[  \lambda_{\hat{\alpha}}^{[abc]|} (x)  , \quad \phi^{[abc]|i}(x) \]
transforming as symplectic traceless anti-symmetric tensor of rank three which we will denote as $14'$.
\end{itemize}

In parallel to full $5d$, $N=8$ supergravity we expect the interacting $(4,0)$ superconformal theory to admit  two maximal $(1,0)$ supersymmetric truncations. By discarding the 6 generalized gravitino multiplets and 14 hyper multiplets one  obtains an interacting theory describing the coupling of  non-metric $(1,0)$  graviton supermultiplet to 14 tensor multiplets with the global symmetry group $SU^*(6)$ under which 15 tensor fields, including the gravitensor transform irreducibly.  14 scalar fields belong to the coset $SU^*(6)/USp(6)$  in the quaternionic magical theory in $d=5$. On the other hand by discarding the 6 non-metric gravitino multiplets together with the 14 tensor multiplets one obtains a consistent truncation to a $6d$ theory describing the coupling of  non-metric $(1,0)$ graviton supermultiplet to 14 hypermultiplets  which reduces in five dimensions to $N=2$ supergravity coupled to 14 hypermultiplets with the scalar manifold $F_{4(4)}/USp(6)\times SU(2)$\cite{Gunaydin:1983rk}.

The $(1,0)$ truncation of $(4,0)$ theory that reduces to the quaternionic magical theory in 5d can be further truncated such that the resulting theory describes a unified theory in $d=5$.  First, by restricting to the $U(1)$ invariant sector in the decomposition of $SU^*(6)$ with respect to its subgroup $SL(3,\mathbb{C})\times U(1)$, which requires discarding 6 tensor multiplets, one obtains a theory describing the coupling of non-metric ($1,0$)  graviton supermultiplet to eight tensor multiplets that reduces to the complex magical supergravity theory in $d=5$ whose moduli space is $SL(3,\mathbb{C})/SU(3)$.  Second, by a further restriction to $\mathbb{Z}_2$ invariant sector under the decomposition  $SL(3,\mathbb{C})\supset SL(3,\mathbb{R}) \times \mathbb{Z}_2$ one obtains a 6d theory  describing the coupling to 5 tensor multiplets that reduces to the real magical supergravity in $5d$ with the moduli space $SL(3,\mathbb{R})/SO(3)$.

In all the above truncations of the interacting $(4,0)$ theory to a $(1,0)$ supersymmetric theory describing the coupling of generalized graviton multiplet to tensor multiplets all the tensor fields, including the gravitensor, transform in an irrep of the global symmetry group, which are $SU^*(6), SL(3,\mathbb{C})$ and $SL(3,\mathbb{R})$, respectively.
The quaternionic  theory with global $SU^*(6)$ symmetry can be extended to the octonionic theory with  the global symmetry group $E_{6(-26)}$ symmetry by coupling additional 12 tensor multiplets  and reduces to the octonionic magical  supergravity theory with 27  vector fields in $d=5$.
\subsection{ $(3,0)$ supersymmetric truncation of $(4,0)$  supergravity }
\setcounter{equation}{0}
Using the labelling of indices as in the previous subsection one can show that by discarding two $N=6$ gravitensorino multiplets consisting of the fields
\[ \psi^i_{(\hat{\alpha}\hat{\beta}\hat{\gamma})} (x),  \quad  h^{ia}_{(\hat{\alpha}\hat{\beta})}(x) , \quad \lambda_{\hat{\alpha}}^{i[ab]|} (x) , \quad
\phi^{i[abc]|}(x) \]
one obtains the non-metric $(3,0)$ graviton supermultiplet consisting of the fields
\[  R_{(\hat{\alpha}\hat{\beta} \hat{\gamma}\hat{\delta})}(x),  \quad  \psi_{(\hat{\alpha}\hat{\beta}\hat{\gamma})}^a (x),  \quad  h_{(\hat{\alpha}\hat{\beta})}^{[ab]|}(x)+ h^{[ij]}_{(\ha\hb)}(x) \, , \quad \lambda_\ha^{[abc]|}(x) + \lambda^{[ij]a}_\ha (x) \, , \quad \phi^{[ij][ab]|}(x)  \]
The truncation to the non-metric $(3,0)$ supergravity theory has the same bosonic field content as the truncation to the maximal $(1,0)$ tensor Einstein supergravity with 14 self-dual tensor multiplets and the 14 scalars.
The corresponding result for the $5d$ supergravity, namely that $N=6$ supergravity has the same bosonic content as the quaternionic magical theory with the symmetric target space $SU^*(6)/USp(6)$  was was shown in \cite{Gunaydin:1983rk}.
\section{ ExFT  formulation of metric and non-metric  $(1,0)$ magical supergravity theories in six dimensions\label{ExFT_magical}}
The unified ExFT formulation of the bosonic sector of linearized $(4,0)$  and $(2,2)$  Poincare supergravities as formulated  in \cite{Bertrand:2020nob} can  be readily extended to a unified construction  of $(1,0)$ metric and non-metric supergravity theories  in six dimensions that descend to the four magical supergravity theories in five dimensions.  For the octonionic magical supergravity which can not be obtained from maximal supergravity by truncation  one imposes the  section constraint
\be
C^{IJK} \partial_J \otimes \partial_K - \frac{1}{\sqrt{10} } \Delta^{IJ} ( \partial_J \otimes \partial_\bullet +\partial_\bullet \otimes \partial_J ) =0
\ee
where the  C-tensor $C^{IJK}$   is the one given by the cubic norm of  the real exceptional Jordan algebra $J_3^{\mathbb{O}}$ which is Euclidean. It an invariant tensor of  $E_{6(-26)}$.  To obtain the ExFT formulation of the metric $(1,0)$ supergravity in $d=6$ one first decomposes the indices of 27 dimensional representation of $E_{6(-26)}$ with respect ot its $SO(9,1)$ subgroup
\be
C^{0ab} = \frac{1}{\sqrt{10}} \eta^{ab} \quad , \quad C^{a\alpha\beta} = \frac{1}{2\sqrt{5}} (\Gamma^a)^{\alpha\beta}
\ee
where $\eta^{ab}$ is the $SO(9,1)$ invariant metric and $(\Gamma^a)^{\alpha\beta} $ are the gamma matrices of $SO(9,1)$ with $(a,b,..=0,1,..9)$ and $(\alpha , \beta, ..=1,2,...,16$). Imposing the conditions
\be
\partial_{\alpha}  = \partial_a = \partial_{\bullet} =0
\ee
solves the section constraint and the resulting ExFT describes the metric octonionic magical supergravity in $d=6$ with 9 tensor multiplets and 16 vector multiplets and  scalar manifold $SO(9,1)/SO(9)$ \cite{Gunaydin:2010fi}.

For constructing the  non-metric $(1,0)$ supergravity theory one imposes the condition $\partial_I=0$ and identifies the coordinate $Y^\bullet$ with the spatial singlet component $y$ in the $(5+1)$ split of external $6d$ spacetime coordinates $x^\mu$ as was done for the maximal supergravity in \cite{Bertrand:2020nob}.

To obtain the unified ExFT  formulations of metric and non-metric $(1,0)$ supergravity theories that reduce to the linearized quaternionic, complex and magical supergravity theories in $d=5$ one needs to simply substitute the C-tensors of these theories in the section constraint \ref{gensection}  and decompose the indices with respect to subgroups of their $5d$ U-duality groups listed in the first column of Table \ref{tab:reality} and the gamma matrices by those listed in column 4 of that table.
For the quaternionic magical theory  the tensor $C^{IJK}$ becomes the invariant tensor of the internal Lorentz ( reduced structure) group $SU^*(6)$ of $J_3^{\mathbb{H}}$. The resulting metric $(1,0)$ supergravity describes the coupling of 6 tensor multiplets and eight vector multiplets to metric (1,0) supergravity.   The corresponding non-metric $(1,0)$ supergravity theory describes the coupling of 14 tensor multiplets to non-metric $(1,0)$ supergravity. In contrast to the metric theory the non-metric supergravity theory describes  a unified theory since the 15 tensor multiplets transform irreducibly under the global symmetry group $SU^*(6)$. Since the quaternionic magical supergravity can be embedded both in maximal supergravity and octonionic magical supergravity in five dimensions the corresponding ExFts describing the metric and non-metric $(1,0)$ supergravity theories in $6d$ can also be obtained by truncation of the unified formulation of ExFTs describing $(4,0)$ and $(2,2)$ theories in $6d$.


The ExFT defined by $J_3^{\mathbb{H}}$ can be consistently  truncated to a $(1,0)$ unified non-metric tensor-Einstein supergravity described by the Euclidean Jordan algebra $J_3^{\mathbb{C}}$ describing the coupling of 8 tensor multiplet to non-metric $(1,0)$ supergravity. It is an invariant tensor of the Lorentz ( reduced structure ) group $SL(3,\mathbb{C})$ of $J_3^{\mathbb{C}}$. The latter theory can be further truncated to an ExFT corresponding to the real magical supergravity in $d=5$ defined by $J_3^{\mathbb{R}}$. Its C-tensor is invariant under $SL(3,\mathbb{R})$ and describes the coupling of 5 self-dual tensors to non-metric (1,0) supergravity in six dimensions.

\section{  Unified non-metric Tensor-Einstein Supergravity theories in $d=6$  }
As we reviewed in section \ref{unifiedMESGT}  there exist three infinite families of unified $N=2$ MESGTs
in $d=5$. Their C-tensors are given by the structure constants $d_{IJK}$  of the traceless elements of Lorentzian Jordan algebras $J_{(1,N)}^{\mathbb{A}} $  of degree three over the associative division algebras $\mathbb{A}=\mathbb{R},\mathbb{C},\mathbb{H}$ which are invariant under their automorphism groups of $J_{(1,N)}^{\mathbb{A}} $.
Since the  structure constants $d_{IJK}$ of three Lorentzian Jordan algebras of degree four, namely
$J_{(1,3)}^{\mathbb R}$, $J_{(1,3)}^{\mathbb C}$ and $J_{(1,3)}^{\mathbb H}$,
coincide with the C-tensors defined by the norms of the Euclidean Jordan algebras of degree three,
$J_{3}^{\mathbb C}$, $J_{3}^{\mathbb H}$ and $J_{3}^{\mathbb O}$ respectively\cite{Gunaydin:2003yx}, they have accidental enlarged hidden symmetries:
\bea
Aut [J_{(1,3)}^{\mathbb R}]= SO(3,1) \quad \Longrightarrow \quad Str_0(J_3^{\mathbb{C}})= SL(3,\mathbb{C}) \\
Aut [J_{(1,3)}^{\mathbb C}]= SU(3,1) \quad \Longrightarrow \quad Str_0(J_3^{\mathbb{H}})= SU^*(6) \\
Aut [J_{(1,3)}^{\mathbb H}]= USp(6,2) \quad \Longrightarrow \quad Str_0(J_3^{\mathbb{O}})=E_{6(-26)}
\eea
The enlarged symmetry groups  $SL(3,\mathbb{C}), SU^*(6) $ and $E_{6(-26)}$ are the global symmetry  groups of the $5d$ MESGTs defined by the corresponding Euclidean Jordan algebras.
As we discussed above these three $5d$ unified MESGTs theories can be obtained from unified tensor-Einstein supergravity theories in $6d$. For the non-metic tensor-Einstein supergravity theories to be unified theories all the tensor fields including the gravitensor must transform irreducibly under the global symmetry group.  Remarkably the corresponding irreducible representations of  $SL(3,\mathbb{C}), SU^*(6) $ and $E_{6(-26)}$ remain irreducible under the restriction to the manifest symmetry subgroups  $SO(3,1),SU(3,1)$ and $USp(6,2)$, respectively.

For the other members of the three infinite families of unified MESGTs there is no symmetry enhancement  beyond the  automorphism groups of the underlying Lorentzian Jordan algebras.  Nonetheless we expect them to descend from unified tensor-Einstein supergravity theories in $6d$ in a similar fashion. In addition to the 3 infinite families there exist a unified MESGT in $d=5$  defined by the Lorentzian octonionic Jordan algebra of degree three $J_{(1,2)}^{\mathbb{O}} $.   This isolated theory is also expected to descend  from a unified tensor-Einstein supergravity in $d=6$.  In Tables \ref{unifiedTESGT} and \ref{unifiedMESGT} we give the list of  unified tensor-Einstein supergravity theories in $d=6$, their field content and global symmetry groups, under which all the tensor fields including the gravitensor form an  irrep.

\begin{table}[ht]
\scriptsize
\begin{center}
\begin{displaymath}
\begin{array}{|c|c|c|c|}
\hline
~&~&~&~\\
J &  \textrm{ Aut}(J)\supset K
& \textrm{Decompositon of self-dual tensor fields under $K$} & \textrm{Number  of scalars}  \\
\hline
~&~&~&~\\
J_{(1,N)}^{\mathbb R} & SO(N,1)\supset SO(N) &
\frac{1}{2}N(N+3)=[\frac{1}{2}N(N+1)-1]\oplus N \oplus 1 &\frac{1}{2}N(N+3)-1 \\
~&~&~&~\\
J_{(1,N)}^{\mathbb C}  & SU(N,1)\supset SU(N)\times U(1) & N(N+2)= [N^2-1]^0 \oplus N^{+1} \oplus N^{-1} \oplus 1^0 & N(N+2)-1 \\
~&~&~&~\\
J_{(1,N)}^{\mathbb H}  & USp(2N,2)\supset USp(2N)\times USp(2) & N(2N+3)=[(N(2N-1)-1,1)]\oplus (2N,2) +(1,1)  & N(2N+3)-1 \\
~&~&~&~\\
J_{(1,2)}^{\mathbb O} & F_{4(-20)}\supset SO(9) & 26=9 \oplus 16 \oplus 1 & 25 \\
~&~&~&~ \\
\hline
\end{array}
\end{displaymath}
\end{center}

\caption{\label{unifiedTESGT} Unified tensor-Einstein supergravity theories in $d=6$ that reduce to unified MESGTs in $d=5$ defined by simple Lorentzian Jordan algebras
$J_{(1,N)}^{\mathbb{A}}$ whose scalar manifolds are not homogeneous spaces for $N\neq 3$. Second  column lists
  their
global symmetry  groups $\textrm{Aut}
(J_{(1,N)}^{\mathbb{A}})$ and their maximal compact subgroups $K$. Third  column lists
dimension of the irrep of tensor fields under $\textrm{Aut}
(J_{(1,N)}^{\mathbb{A}})$   and their decomposition with respect to the maximal compact subgroup $K$. The last column lists the number of scalars. The theories with $N=3$ have hidden larger symmetries and correspond to three of the magical supergravity theories. }
\end{table}
\begin{table}[ht]
\scriptsize
\begin{center}
\begin{displaymath}
\begin{array}{|c|c|c|c|}
\hline
~&~&~&~\\
J &  \textrm{Global symmetry group}
& \textrm{Number of self-dual tensor fields} & \textrm{Scalar Manifold}  \\
\hline
~&~&~&~\\
J_{3}^{\mathbb R} & SL(3,\mathbb{R})\supset SO(3) &
6= 5 \oplus 1 & SL(3,\mathbb{R})/SO(3) \\
~&~&~&~\\
J_{3}^{\mathbb C}  & SL(3,\mathbb{C})\supset SU(3) & 9= 8  \oplus 1  & SL(3,\mathbb{C})/SU(3) \\
~&~&~&~\\
J_{3}^{\mathbb H}  & SU^*(6)\supset USp(6) & 15=14 \oplus 1  & SU^*(6)/USp(6) \\
~&~&~&~\\
J_{(3)}^{\mathbb O} & E_{6(-26)} \supset F_4 & 27=26 \oplus 1 & E_{6(-26)}/F_4 \\
~&~&~&~ \\
\hline
\end{array}
\end{displaymath}
\end{center}

\caption{\label{unifiedMESGT} Unified tensor-Einstein supergravity theories in $d=6$ that reduce to magical supergravity theories in $d=5$ defined by simple Euclidean Jordan algebras $J_{3}^{\mathbb{A}}$ of degree three where $\mathbb{A}=\mathbb{R},\mathbb{C},\mathbb{H},\mathbb{O}$.
 Second  column lists
  their
global symmetry  groups  and their maximal compact subgroups. Third  column lists
the number of tensor fields  and their decomposition with respect to the maximal compact subgroup. The last column lists their scalar manifolds which are symmetric spaces. }
\end{table}

As is evident from the Tables \ref{unifiedTESGT} and \ref{unifiedMESGT} in the decomposition of the irreducible representation of the global symmetry group with respect to its maximal compact subgroup there is a unique singlet which is to be identified with  the "bare gravitensor".
Furthermore all the bosonic fields in tensor Einstein supergravity  theories are singlets of the $R$-symmetry group $USp(2)$.

 One can gauge certain subgroup of the global symmetry groups of the unified MESGTs in $d=5$. This naturally leads to the question whether  the corresponding six dimensional tensor-Einstein supergravity theories admit interactions among tensor fields that reduce to the non-Abelian gauge interactions in five dimensions. It is generally believed that the interacting superconformal $(2,0)$ theories do not admit a Lagrangian formulation since they decribe multiple M-5 branes that are strongly coupled with no free parameter for formulating a perturbative Lagrangian theory. Nonetheless a novel method of introducing such non-Abelian couplings in certain $(1,0)$ superconformal field theories in $d=6$ was developed  by Samtleben and collaborators \cite{Samtleben:2011fj,Samtleben:2012fb,Samtleben:2012mi}.  It is an open problem whether some of these theories can be coupled to non-metric
$(1,0)$ supergravity in $6d$ that upon dimensional reduction reduce to $5d$, $N=2$ Yang-Mills-Einstein supergravity theories.

\subsection{Dimensional reduction of non-metric $(4,0)$ and $(1,0)$ tensor Einstein supergravity theories}

Dimensional reduction of $(4,0)$ supergravity multiplet using vectorial indices was performed by Hull who showed that the resulting field content coincides with that of maximal Poincare supergravity in five dimensions. Dimensional reduction of the $(4,0)$ unitary supermultiplet using the twistorial spinor indices was given in \cite{Chiodaroli:2011pp}. Twistorial oscillator method yields manifestly unitary supermultiplets which involve only the physical degrees of freedom. Hence the resulting supermultiplets involve only the field strengths and not the gauge fields. For the doubleton
supermultiplet of $OSp(8^*|8)$ of the $(4,0)$ non-metric supergravity in $d=6$ the field strengths
correspond to symmetric tensors in the spinor indices $\ha,\hb, ..$ of the Lorentz group $SU^*(4)$.
Since the spinor representation $S_\ha$  of 6 dimensional Lorentz group $SU^*(4)$ and symmetric tensor representations $ S_{\ha\hb\hd,...}$ remain irreducible  under the restriction to the five dimensional Lorentz group $USp(2,2)$, the dimensional reduction to $d=5$  in the twistorial formulation as given in \cite{Chiodaroli:2011pp} is much simpler than in the formulation involving vectorial indices. There is a one-to-one correspondence between the field strengths belonging to the doubleton supermutiplet of $OSp(8^*|8)$ and the field strengths of the fields of maximal supergravity in $d=5$. In particular non-metric graviton in $6d$ with the field strength $R_{(\ha\hb\hd\hg)}$ reduces to the five dimensional graviton field strength without an extra vector or scalar field\cite{Hull:2000rr,Chiodaroli:2011pp}. This is to be contrasted with the $6d$ metric graviton which reduces to a graviton plus a vector and a scalar. Similarly the gravitensor field strength $\psi^A_{(\ha\hb\hg)}$ remains irreducible under restriction to $USp(2,2)$ subgroup of $SU^*(4)$ and becomes the field strength of a gravitino in $5d$:
\[ \psi^A_{(\alpha\beta\gamma)}= \partial_{[\mu} \psi^A_{\nu] (\alpha} *\gamma^{\mu\nu}_{\beta\gamma)} \]
where $*\gamma^{\mu\nu}=\frac{1}{3!}\epsilon^{\mu\nu\lambda\epsilon\delta} \gamma_{\lambda\epsilon\delta}$.
Self-dual tensor fields reduce to  vector fields in $d=5$ and the symplectic Majorana Weyl spinors go over to symplectic Majorana spinors in $d=5$.

Therefore non-metric $(1,0)$ supersymmetric tensor Einstein supergravity theory in $d=6$ will reduce to a five dimensional $N=2$  supergravity with the same number of $5d$ vector fields as self-dual tensors in $6d$. The bare gravitensor in $6d$ will reduce to the bare graviphoton in $5d$.  What distinguishes unified tensor-Einstein supergravity theories from others is the fact that gravity sector can not be decoupled from the tensorial matter sector without breaking their global symmetry groups since the gravitensor together with the other tensor fields transform irreducibly under them. Ungauged tensor Einstein theories reduce to Maxwell-Einstein supergravity theories in $d=5$.

 Unified MESGT theories, in particular the magical supergravity theories , admit gaugings with simple gauge groups with or without tensor fields in five dimensions. It was shown in \cite{Gunaydin:2010fi} that Poincare uplifts of magical supergravity theories to six dimensions do not admit gaugings with simple groups. Furthermore Poincare uplifts of magical supergravity theories in $6d$ are no longer unified since some of the vector fields uplift to selfdual tensors while the others uplift to vector fields in $6d$.  In addition one finds that magical Poincare supergravity theories in $6d$ admit a unique gauge group which is a nilpotent Abelian group with $(n_T-1)$ translation generators, where $n_T$ is the number of tensor multiplets coupled to $(1,0)$ metric supergravity. These $(n_t-1)$ generators can not lie within the isometry group $SO(n_T,1)$ of the scalar manifold due to appearance of central extensions of the gauge algebra. The gauge algebra with the central extension can be embedded within the isometry group of the corresponding five dimensional magical supergravity. On the other hand in the uplift of the magical supergravity theories to  six dimensions as non-metric $(1,0)$ tensor Einstein supergravity the isometry group of the scalar manifold of the five dimensional theory becomes a global symmetry group in $6d$.

 \subsection{ On the general ExFT formulation of unified non-metric tensor-Einstein supergravity theories in six dimensions.}

 The 27 dimensional internal space-time that is intoduced in ExFT formulation of the $5d$ maximal Poincare supergravity can be identified with the generalized spacetime coordinatized by the split exceptional Jordan algebra $J_3^{\mathbb{O}_s}$ of $3\times 3$ Hermitian matrices over the split octonions $O_s$. This generalized  space-time was first intoduced in the early days of spacetime supersymmetry before any supergravity theories was written down \cite{Gunaydin:1975mp}. Its automorphism, reduced structure and linear fractional groups were identified with the rotation, Lorentz and conformal groups of this space-time which are $F_{4(4)}, E_{6(6)}$ and $E_{7(7)}$ respectively. The generalized space-times defined by Jordan algebras were later studied further in \cite{Gunaydin:1979df,Gunaydin:1992zh,Gunaydin:2005zz,Mack:2004pv}. For the maximal supergravity in $d=5$ the Lorentz group $E_{6(6)}$ is the invariance of the C-tensor as well as of the Lagrangian.

 For the $5d$ octonionic magical supergravity theory the internal space-time is the generalized spacetime defined by the real exceptional Jordan algebra $J_3^{\mathbb{O}}$ of $3\times 3$ Hermitian matrices over the division algebra of octonions whose rotation, Lorentz and conformal  groups are $F_4$,  $E_{6(-26)}$ and $E_{7(-25)}$, respectively.  The invariance groups of the C-tensors of magical supergravity theories are given by the Lorentz groups of underlying Euclidean Jordan algebras of degree three. As was discussed above the complex, quaternionic and octonionic magical supergravity theories can equivalently be described by the Lorentzian Jordan algebras of  degree four over the reals $\mathbb{R}$ , complex numbers $\mathbb{C}$ and quaternions $\mathbb{H}$. The C-tensor $C^{IJK}$ given by the cubic norm of $J_3^{\mathbb{C}},J_3^{\mathbb{H}}$ and $J_3^{\mathbb{O}}$ coincides with the structure constants of the traceless elements of the Lorentzian Jordan algebra degree four $J_{(1,3)}^{\mathbb{R}}$ , $J_{(1,3)}^{\mathbb{C}}$ and $J_{(1,3)}^{\mathbb{H}}$, respectively.
 As such they belong to three infinite families of unified MESGTs in $5d$ defined by Lorentzian Jordan algebras $J_{(1,n)}^{\mathbb{A}}$ of arbitrary degree  over the reals $\mathbb{R}$ , complex numbers $\mathbb{C}$ and quaternions $\mathbb{H}$.

The unified ExFT formulation of linearized $6d$, $(4,0), (3,1)$ and standard maximal $(2,2)$ supergravity all reduce
to the linearized maximal supergravity in $d=5$\cite{Bertrand:2020nob}. At the linearized level only the maximal compact symmetry group $USp(8)$ of $E_{6(6)}$ is manifest in five dimensions. The maximal compact symmetries groups of $(4,0), (3,1) $ and $(2,2)$ theories in $d=6$ are $USp(8), USp(6)\times USp(2)$ and $USp(4)\times USp(4)$, respectively. The fact that they all lead to the same linearized theory in five dimensions and the standard maximal supergravity theory has a non-linear extension in both five and six dimensions suggests   that the generalized ExFT formulation of the maximal $(2,2)$ supergravity theory in $d=6$ could be extended to a nonlinear interacting $(4,0)$ and $(3,1)$ non-metric supergravity theories in six dimensions as was argued in \cite{Bertrand:2020nob}.

The maximal  $(2,2)$ supergravity theory in six dimensions can be truncated to $(1,0)$ supersymmetric Poincare supergravity that reduces to the magical supergravity defined by the Euclidean Jordan algebra $J_3^{\mathbb{H}}$ in five dimensions. Its global symmetry group in six dimensions $SU^*(4)\times USp(2)$ which is a subgroup of the$5d$ global symmetry group $SU^*(6)$.
\be
SO(5,1)\times USp(2) \subset SO(5,1)\times SO(4) =SU^*(4) \times USp(2)\times USp(2) \subset SO(5,5)
\ee
and describes the coupling of 5 self-dual and 8 vector multiplets to $(1,0)$ metric supergravity. The global symmetry group of this theory gets enlarged to $SU^*(6)$ in five dimensions with the scalars parametrizing the symmetric space $SU^*(6)/USp(6)$. This theory can be further truncated to the  complex and real magical supergravity theories.
The dimensional reduction of unified ExFT formulation of linearized metric and non-metric magical supergravity theories to five dimensions the resulting theories are invariant  only under the maximal compact subgroups of their global symmetry groups which are $F_4, USp(6), SU(3)$ and $SO(3)$.

As summarized in section \ref{unified} $5d$  MESGTs described by the Euclidean Jordan algebras $J_3^{\mathbb{H}}$ and $J_3^{\mathbb{C}}$ can be equivalently formulated using the Lorentzian Jordan algebras $J_{(1,3)}^{\mathbb{C}}$ and $J_{(1,3)}^{\mathbb{R}}$, respectively.   Since  the section constraint for a unified ExFT formulation of these theories depends only on the C-tensor it extends to the formulation in terms of Lorentzian Jordan algebras.
Interestingly the extra singlet coordinate $Y^\bullet$ in the ExFT formalism can now be identified with the identity element of the corresponding Lorentzian Jordan algebra.
Therefore for the three infinite families as well as the sporadic unified non-metric $(1,0)$ TESGTs in $d=6$ one  can give an ExFT formulation of their linearized bosonic sectors using the section constraint
\be
C^{IJK} \partial_J \otimes \partial_K - \frac{1}{\sqrt{10} } \Delta^{IJ} ( \partial_J \otimes \partial_\bullet +\partial_\bullet \otimes \partial_J ) =0  \nonumber
\ee
where $C^{IJK}$ are now the structure constants  of the underlying Lorentzian Jordan algebras which are  invariant tensors of their global symmetry groups given by their automorphism groups.
In all cases the extra singlet coordinate $Y^\bullet$ can be identified with the identity element of the underlying Lorentzian Jordan algebra. Imposing the condition $\partial_I=0$ solves the section constraint trivially and leads to the bosonic sector of  non-metric $(1,0)$ TESGTs in $d=6$.

Apart from the magical supergravity theories the higher dimensional origins of the three infinite families of $5d$ unified MESGTs as Poincare supergravity theories is not known. Whether the  unified section constraint admits solutions that lead to three infinite families of $(1,0)$ supersymmetric metric Poincare supergravities in six dimensions will be left for future investigations.

\section{Conformal path to higher dimensions and non-metric supergravity theories}

If an interacting non-metric $(4,0)$ supergravity exists in $d=6$ that reduces to the maximal supergravity in lower dimensions it raises the question as to whether $d=6$ is the maximal dimension for the existence of non-metric supergravity theories. Now six is the maximal dimension for the existence of conformal superalgebras that extend the conformal algebra of
$SO(d,2)$ to a simple conformal  superalgebra \cite{Nahm:1977tg}. Here the isomorphism of $SO(6,2)$ to $SO^*(8)$ plays a key role for satisfying spin and statistics constraints. The extensions of  the Lie algebra of $SO(d,2)$ to simple Lie superalgebras for $d>6$  do not satisfy the correct spin and statistics connection. Therefore it is believed that conformal metric supergravity theories  based on simple superconformal algebras exist only in $d\leq 6$.

 On the other hand Poincare superalgebras, which are not simple, exist in any dimension. However maximal dimension for Poincare supergravity is $d=11$ \cite{Nahm:1977tg}. The Poincare superalgebra in $d=11$ with 32 supercharges can be embedded in the simple Lie superalgebra $OSp(1|32,\mathbb{R})$ with the even subalgebra $Sp(32,\mathbb{R})$ \cite{vanHolten:1982mx,Townsend:1997wg,Gunaydin:1998bt,Bergshoeff:2000qu}.
Its contraction to 11 dimensional Poincare superalgebra involves tensorial central charges
\eq
\{ Q_{A}, Q_{B} \} = (C\Gamma^M)_{AB} P_M + \frac{1}{2}
(C\Gamma^{MN})_{AB} Z_{MN} + \frac{1}{5!} (C\Gamma^{M_1..M_5})_{AB}
Y_{M_1 ..M_5}  \label{eq:malgebra}
\en
where $\Gamma^M  (M,N=0,1,..,10)$ are the  Dirac gamma matrices and $C$ is the charge conjugation matrix.
The embedding is such that the fundamental representation of $Sp(32,\mathbb{R})$ is identified with the Majorana spinor of the Lorentz group $SO(10,1)$. The $Z_{MN}$ and $Y_{M_1,...,M_5}$ are the tensorial central charges  with $P_M$ representing translations.

 Werner Nahm's classification of spacetime superalgebras assumed that the spacetime has Minkowskian signature and gravity is described by a spacetime metric or a corresponding spin connection. In the $(4,0)$ supermultiplet of $OSp(8^*|8)$ the gauge field of the graviton field strength is not a spacetime metric but rather a mixed tensor both in the first order as well as the second order formalism as reviewed above. As was shown by Hull the gauge symmetries of the mixed tensor reduce to diffeomorphisms in five dimensions and the interacting theory should yield the standard maximal supergravity in five dimensions. This raises the question whether there could be higher dimensional non-metric supergravity theories which reduce to the interacting $(4,0)$ theory in $d=6$ or standard  maximal supergravity in five and lower dimensions. At the level of Lie superalgebras the answer appears to be yes. Namely there exist superalgebras of the form $OSp(2n^*|2m)$ which extend the Lie superalgebra $OSp(8^*|8)$ to higher dimensions. For 64 supercharges the possibilities are $OSp(16^*|4)$ and $OSp(32^*|2)$.

 In this context we should point out that M-theory was studied  in space-times with exotic  signatures by Hull \cite{Hull:1998ym}. Later M-theory and superstring theory in exotic signature space-times were studied within the framework of negative branes in \cite{Dijkgraaf:2016lym}. Typically the Lorentz groups of these exotic spacetimes are of the form $SO(p,11-p)$ or $SO(p-10-p)$ for formulations of  M-theory and IIB superstring theory, respectively and the corresponding gravitational theories are of metric type.  To our knowledge formulation of M/superstring theory on exotic spacetimes with conformal groups of type $SO^*(2n)$ that admit interpretation as non-metric gravity theories  has not yet been investigated.

  One natural framework for going beyond standard Minkowskian spacetimes with metric gravity is that of generalized spacetimes coordinatized by Euclidean Jordan algebras\cite{Gunaydin:1975mp,Gunaydin:1989dq,Gunaydin:1992zh,Gunaydin:2005zz}. The conformal groups of space-times defined by Euclidean Jordan algebras all admit positive energy unitary representations \cite{Gunaydin:1992zh} and were shown to be causal space-times in \cite{Mack:2004pv}. The standart critical Minkowskian space-times that admit supersymmetric Yang-Mills theories can be coordinatized by Euclidean Jordan algebras $J_2^{\mathbb{A}}$ of degree two generated by Hermitian $2\times 2$ matrices over the four division algebras $\mathbb{A}=\mathbb{R},\mathbb{C},\mathbb{H},\mathbb{O}$. Their conformal groups are $Sp(4,\mathbb{R}), SU(2,2), SO^*(8)$ and $SO(10,2)$, respectively. Except for the octonionic case they all admit  extensions to simple Lie superalgebras, namely $OSp(n|4,\mathbb{R}), SU(2,2|n) $ and $ OSp(8^*|2n)$ which all satisfy the usual spin and statistics connection.\footnote{ The octonionic conformal algebra $SO(10,2)$ has an extension to the simple Lie superalgebra $OSp(1|32,\mathbb{R})$ which requires its extensions to the Lie algebra $Sp(32,\mathbb{R})$.}  The natural extension of these space-times is to consider those defined by Euclidean Jordan algebras $J_3^{\mathbb{A}}$ of degree 3 over the four division algebras $\mathbb{A}=\mathbb{R},\mathbb{C},\mathbb{H},\mathbb{O}$ which were studied in \cite{Gunaydin:2005zz}.  These  spacetimes correspond to extensions of the Minkowskian space-times by twistorial coordinates  given in the second column of table \ref{tab:reality} and an extra singlet coordinate.  The C-tensors given by the norm forms of Jordan algebras of degree three satisfy the so-called adjoint identity
 \begin{equation}
   C^{IJK} C_{J(MN} C_{PQ)K} = {\delta^I}_{(M} C_{NPQ)}
\end{equation}
 It is a remarkable but little known fact that for simple Jordan algebras of degree three the adjoint identity implies the Fierz identities for the existence of supersymmetric Yang-Mills theories in the critical dimensions\cite{Sierra:1986dx}.
 The conformal groups of $J_3^{\mathbb{A}}$ are $Sp(6,\mathbb{R}), SU(3,3), SO^*(12)$ and $E_{7(-25)}$, respectively. Again except for the octonionic case they admit extensions to simple superalgebras $OSp(n|6,\mathbb{R}), SU(3,3|n)$ and $OSp(12^*|2n)$, respectively.
The conformal superalgebras in critical dimensions $3,4$ and 6 are subalgebras of these superalgebras.

 We should note that $SO^*(4n)$ is the conformal group of a spacetime coordinatized by the Euclidean Jordan algebra $J_n^{\mathbb{H}}$ of $ n \times n $ Hermitian matrices over the division algebra of quaternions \cite{Gunaydin:1992zh}. Its Lorentz and rotation groups are $SU^*(2n)$ and $USp(2n)$, respectively.  The quaternionic Jordan algebra  $J_2^{\mathbb{H}}$ of degree two describes the standard $6d$ Minkowskian space-time with the Lorentz group $SU^*(4)$ and conformal group $SO^*(8)$ , which are isomorphic to $Spin(5,1)$ and $SO(6,2)$, respectively.  The conformal superalgebras $OSp(8^*|2n)$ come in three versions due to triality. The usual spin and statistics connection requires the version for which the supersymmetry generators transform in the spinor representation of $SO(6,2)$ which decomposes as $ (4 + \bar{4})$ with respect to the Lorentz subgroup $SU^*(4)$. The six dimensional Minkowski coordinates are anti-symmetric tensors $x^{\alpha\beta}=-x^{\beta\alpha}$ in spinorial indices $\alpha, \beta$ as discussed earlier.  For $n>2$ the supersymmetry generators of $OSp(4n^*|2m)$ decompose as $(2n) + \overline{(2n)}$ with respect to the "Lorentz " group $SU^*(2n)$ . The 2n-dimensional representation of $SU^*(2n)$ is the analog of spinor representation of $SU^*(4)$ and may be interpreted as generalized twistorial coordinates.\footnote{The corresponding generalized twistorial coordinates for the space-times defined by complex Jordan algebras $J_n^{\mathbb{C}}$  appeared  in the work of \cite{Brody:2004jk} without using the language of Jordan algebras as generalizations of the twistor theory of Penrose and were called hyperspinors.}

If indeed there exist non-metric supergravity theories that reduce to standard metric supergravity theories one would have to extend  the analysis of Werner Nahm with regard to classification of physically relevant spacetime superalgebras.
One possible way to look for such exotic supergravity theories is to study solutions of ExFT formulation of maximal supergravity in 5,4 and 3 dimensions that admit interpretation as non-metric supergravity theories in  higher dimensions.
   Investigations of these questions and their implications  for  M/superstring theory will be left to future investigations.

{\bf Note added to the revised version: } The results of the first version of this paper were obtained before the work of \cite{Bertrand:2020nob} and \cite{Minasian:2020vxn}  on $(4,0)$ supergravity in $6d$ appeared in the arXiv. The authors of  \cite{Minasian:2020vxn} study  the anomalies in $(4,0)$ supergravity and its truncations  and argue that only their dimensional reduction to three dimensions make contact with standard metric supergravity theories.  In \cite{Bertrand:2020nob} the free $(4,0)$ theory is studied from the point of view of exceptional field theory. In this revised version we  review briefly the work of \cite{Bertrand:2020nob} on ExFT formulation of linearized $(4,0)$ theory ( subsection 5.2) and extend it  to ExFT formulation of linearized unified $(1,0)$
tensor Einstein supergravity theories in $6d$ ( section 7).

 {\bf Acknowledgements:} I would like to thank  Stanford Institute of Theoretical Physics for its kind hospitality where this work was carried out.  I would  like to thank Marc Henneaux for posing a question that triggered the invesigation that led to this paper and   Henning Samtleben for a discussion on an earlier draft of this paper. Thanks  also to Marco Chiodaroli, Henrik Johansson  and  Radu Roiban for their helpful comments on the results reported above. I would also like to thank the referee for his/her insightful comments and helpful suggestions.

 \appendix

\section{CPT-self-conjugate  Supermultiplet of $SU(2,2|8)$ }
In  \cite{Gunaydin:1984vz} it was shown that the fields of maximal $N=8$ supergravity can be fitted into the CPT-self-conjugate doubleton supermultiplet of the conformal superalgebra $SU(2,2|8)$ with 64 supersymmetry generators. We reproduce this supermultiplet  in Table \ref{Table_GM}. \\

\begin{table}[ht]
\begin{center}
\begin{tabular}{|c|c|c|c|c|}
\hline
   ${ SU(2)_{j_1}\times SU(2)_{j_2} } $ & ${ E_0} $ & $ {  SU(8)}$ & ${ U(1)_R}$ &{ Fields}
\\ \hline
 $(0,0)$ & 1     & ${\bf 70}$   & 0 &$\phi^{[ABCD]} $
\\ \hline
 $({1 \over 2},0)$ &  ${3 \over 2} $& ${ \bf 56}$ & 1/2 &$\lambda^{[ABC]}_{+} \equiv \lambda_{\alpha}^{[ABC]}$
\\ \hline
 $(0,{1 \over 2})$   & ${3 \over 2}$   &$\overline{ \bf 56}$ & -1/2 &$\lambda_{-[ABC]} \equiv \lambda_{\dot{\alpha}[ABC]}$
\\ \hline
 (1,0)  & 2 & ${\bf 28}$& 1 &$h_{\mu\nu}^{+[AB]} \equiv h_{(\alpha\beta)}^{[AB]}$
\\ \hline
(0,1)  & 2 & $\overline{\bf 28}$& -1 &$h_{\mu\nu [AB]}^{-} \equiv h_{(\dot{\alpha}\dot{\beta})[AB]}$
\\ \hline
 $({3 \over 2},0)$  & ${5 \over 2}$ & ${\bf 8}$& 3/2 &$\partial_{[ \mu}\psi_{\nu]}^{+ A} \equiv \psi_{(\alpha\beta\gamma)}^{A}$
\\ \hline
 $(0,{3 \over 2})$  & ${5 \over 2}$ & $\bar{\bf 8}$& -3/2 &$\partial_{[ \mu}\psi_{\nu] A}^{-} \equiv \psi_{(\dot{\alpha}\dot{\beta}\dot{\gamma}) A}$
\\ \hline
 $(2,0)$  &  3 & $1 $& 2 &$ R_{(\alpha\beta\gamma\delta)}$
\\ \hline
 $(0,2)$  & 3 & $1 $& -2 &$ R_{(\dot{\alpha}\dot{\beta}\dot{\gamma}\dot{\delta} )}$
\\ \hline
\end{tabular}
\medskip
\caption{\small \label{Table_GM}
The CPT-self-conjugate doubleton supermultiplet of the superconformal algebra $SU(2,2|8)$
The first two columns list the  $SU(2)_{j_1} \times SU(2)_{j_2}$ and $U(1)_E$ labels of the positive energy unitary representation of $SU(2,2)$ with respect to its maximal compact subgroup. Third column lists their $SU(8)_R$ transformations and fourth column lists the $U(1)_R$ labels corresponding  to their helicities. The last column lists the Lorentz covariant fields  as well as their $SU(8)$ labelling.
The indices $A,B,C,..=1,2,..,8 $ are the SU(8) $R$-symmetry  indices. The indices $\alpha, \beta,..$ and $\da, \db, ..$ denote the left-handed and right-handed spinorial indices of $SL(2,\mathbb{C})$. Round (square) brackets denote complete symmetrization (antisymmetrization) of  indices.}
\end{center}
\end{table}
In the full  nonlinear  ${\cal N}=8$ supergravity superconformal symmetry $SU(2,2|8)$ is broken down  to ${\cal N}=8$ Poincar\'e supersymmetry. Whether an interacting superconformal theory based on the supermultiplet of Table \ref{Table_GM} exists was posed as an open problem in \cite{Gunaydin:1984vz}.
To date no conformal supergravity with the same field content as maximal $N=8$ supergravity in four dimensions has been constructed. It is clear that such a supergravity would have to have some unusual properties such as non-locality and higher derivaties.
On the other hand the superalgebra $SU(2,2|8)$ and the above  supermultiplet
play a key role  in the construction and classification  of potential higher-loop counterterms in maximal
supergravity \cite{Beisert:2010jx}.

The  physical fields of the CPT-self-conjugate doubleton supermultiplet of $SU(2,2|8)$ can be  represented as a scalar
superfield $W_{abcd}$ \cite{Siegel:1980bp,Howe:1981qj,Chiodaroli:2011pp} in the superspace with coordinates,
\be \big( x^{\a \dot \b}, \theta^{\a a} , \bar \theta^{\dot \b}_b   \big) \ee
where $\a,\dot \b= 1,2 $ and $a=1,2,\dots 8$. The  covariant derivatives in this superspace satisfy
\be \{ D_{\a a}, \bar{D}_{\dot \b}^b \} = 2 i \delta_{a}^b \partial^{\vphantom{A}}_{\a \dot \b} \ .
\label{comm-D-4d}  \ee
The superfield $W_{abcd}$  is completely anti-symmetric and obeys the self-duality condition
\be \bar{W}^{abcd} = {1 \over 4 !} \epsilon^{abcdefgh} W_{efgh} \ee
as well as the  differential constraint
\be \bar{D}_{\dot \a}^a W_{bcde} + {4 \over 5} \delta^a_{[b} \bar{D}^f_{\dot  \a} W_{cde]f} = 0 \ . \label{diffcon4D} \ee

\providecommand{\href}[2]{#2}\begingroup\raggedright\endgroup

\end{document}